\documentclass[preprint]{aastex}
\include{epsf}
\usepackage{longtable}

\def\Msun{\hbox{$\rm\thinspace M_{\odot}$}}
\def\Zsun{\hbox{$\rm\thinspace Z_{\odot}$}}

\shorttitle{Galactic and M31 Globular Clusters : I}
\shortauthors{Beasley et al.}

\begin{document}


\title{The Chemical Properties of Milky Way and M31 Globular Clusters: I. A Comparative Study}


\author{Michael A. Beasley}
\affil{Lick Observatory, University of California, Santa Cruz, CA 95064, USA}
\email{mbeasley@ucolick.org}

\author{Jean P. Brodie}
\affil{Lick Observatory, University of California, Santa Cruz, CA 95064, USA}
\email{brodie@ucolick.org}

\author{Jay Strader}
\affil{Lick Observatory, University of California, Santa Cruz, CA 95064, USA}
\email{strader@ucolick.org}

\author{Duncan A. Forbes}
\affil{Centre for Astrophysics \& Supercomputing, Swinburne University,
  Hawthorn, VIC 3122, Australia}
\email{dforbes@astro.swin.edu.au}

\author{Robert N. Proctor}
\affil{Centre for Astrophysics \& Supercomputing, Swinburne University,
  Hawthorn, VIC 3122, Australia}
\email{rproctor@astro.swin.edu.au}

\author{Pauline Barmby}
\affil{Harvard-Smithsonian Center for Astrophysics, 60 Garden Street, 
Cambridge, MA 02138, USA}
\email{pbarmby@cfa.harvard.edu}

%\and

\author{John P. Huchra}
\affil{Harvard-Smithsonian Center for Astrophysics, 60 Garden Street,
Cambridge, MA 02138, USA}
\email{huchra@cfa.harvard.edu}


\begin{abstract}
A comparative analysis is performed between 
high-quality integrated spectral indices of 30 globular clusters
in M31, 20 Milky Way globular clusters, 
and a sample of field and cluster elliptical galaxies.
We find that the Lick CN indices in the M31 and Galactic 
clusters are enhanced
relative to the bulges of the Milky Way, M31, and elliptical
spheroids, in agreement with Burstein et al.
Although not particularly evident in the Lick CN indices, 
the near-UV cyanogen feature ($\lambda3883$)
is strongly enhanced with respect to the Galactic
globulars at metallicities, --1.5$<$[Fe/H]$<$--0.3.
Carbon shows signs of varying amongst these two groups.
For [Fe/H]$>$--0.8, we observe no systematic differences in 
the H$\delta$, H$\gamma$, or H$\beta$ indices 
between the M31 and Galactic globulars, in contrast
to previous studies.
The elliptical galaxy sample lies offset from the
loci of the globular clusters in the both the 
Cyanogen--[MgFe], and Balmer-line--[MgFe] planes.
Six of the M31 clusters appear young, and are
projected onto the M31 disk. 
Population synthesis models 
suggest that these are metal-rich clusters with 
ages 100--800 Myr, metallicities --0.20 $\leq$ [Fe/H] $\leq$
0.35, and masses $0.7\sim7.0\times10^4\Msun$.
Two other young clusters are Hubble V in NGC~205, 
observed as a template, and an older ($\sim$3 Gyr) cluster 
some 7 kpc away from the plane of the disk.
The six clusters projected onto the disk 
show signs of rotation similar to the H{\small I} gas
in M31, and three clusters exhibit thin disk kinematics
(Morrison et al.). Dynamical mass estimates and detailed structural parameters
are required for these objects to determine whether they 
are massive open clusters or globular clusters.
If they are the latter, our findings suggest globular clusters 
may trace the build
up of galaxy disks. In either case, we conclude that these clusters are
part of a young, metal-rich disk cluster system in M31, possibly as 
young as 1 Gyr old.

\end{abstract}



\section{Introduction}

Understanding the formation and evolution of disk galaxies
is a crucial aspect of galaxy formation studies
(Freeman \& Bland-Hawthorn 2002).
Much of our knowledge about such disks
is based upon detailed observations of the Milky Way, and
in this regard, its globular clusters (GCs) have played a
prominent role.
The Milky Way possesses a retinue of $\sim$150 
GCs which, while a minor contributer to 
the total mass budget of the Galaxy 
(they constitute some $\sim$3$\times$10$^7\Msun$ in total, or $\sim$0.1\%
of the baryonic mass of the Milky Way), 
have proved invaluable for understanding its structure, 
stellar populations, and mass aggregation
history (e.g., Searle \& Zinn 1978).
They comprise of at least two components, 
a spatially extended, metal-poor, pressure-supported system, 
and a metal-rich, centrally concentrated system which 
exhibits some rotation (Zinn 1985). Although initially
thought to be associated with the thick disk (Zinn 1985), 
the majority of these metal-rich clusters are now believed to be
associated with the bulge (Frenk \& White 1982; Minitti 1995; 
Cot$\acute{e}$ 1999; Forbes, Brodie \& Larsen 2001).

To date, all the GCs in the Galaxy for which detailed 
color-magnitude diagrams have been obtained appear 'old' (i.e., $>$8 Gyr), 
although a subset of metal-rich GCs may be up to $\sim$20\% younger 
than the inner-halo GCs (Rosenberg et al. 1999; Salaris \& Weiss 2002).
There is also evidence for a spread of ages
among the outer-halo GCs, suggesting either  
a clumpy collapse (Searle \& Zinn 1978) or accreted origin 
(e.g., Cot$\acute{e}$ et al. 2000; Bellazzini et al. 2003). 
The significance of such accretions are currently the 
subject of much debate in light of the presently
favored hierarchical clustering paradigm
(Brook et al. 2003; Martin et al. 2004; 
Forbes, Strader \& Brodie 2004).

Notwithstanding our increasingly detailed knowledge 
regarding the Milky Way
GC system, surprisingly little is known about the 
GC systems of spiral galaxies in general. 
This is largely a result of the difficulties involved
in observing what are relatively poor GC systems (compared
with luminous ellipticals), that are often projected 
against an inhomogeneous background.
However, the gaps in our knowledge about the GC systems
of external spirals is slowly changing, 
with increasingly detailed photometric
(e.g., Goudfrooij et al. 2003; Rhode 2003; Chandar, Whitmore \& Lee 2004) 
and spectroscopic studies
(Chandar, Bianchi \& Ford 1999; Schroder et al. 2002; 
Olsen et al. 2004).

In this regard, our close companion, the Andromeda spiral 
(M31; D$\sim$780 kpc, Holland 1998; Stanek \& Garnavich 1998)
provides an important test-bed for the generality of any Milky Way 
formation theories. It is of similar Hubble type (Sb) to the 
Milky Way, although it is possibly less massive (Cot$\acute{e}$ et al. 
2000; Gottesman, Hunter \& Boonysait 2002).
Unlike the Milky Way, M31 appears to have a predominantly 
metal-rich stellar halo (e.g., Holland, Fahlman \& Richer 1996; 
Bellazini et al. 2003; Rich et al. 2004).

M31 possesses the most populous GC system in the Local Group.
van den Bergh (1999) estimated a total GC system
of 400 $\pm$ 55, which is at least a factor of two larger
than that of the Milky Way.
Previous studies to catalog the M31 GC system have included 
Sargent et al. (1977), Battistini et al. (1980, 1987,
1993) and Crampton et al. (1985). Recently, Barmby et al. (2000) 
published a comprehensive photometric and
spectroscopic catalog of 435 GCs and GC candidates.
These authors showed that the M31 GC system could be divided
into metal-poor and metal-rich sub-populations, 
with their kinematics separating into halo and 
bulge/disk components similar to that of the Milky Way.
Perrett et al. (2002) obtained intermediate resolution spectra for 
a subset of $\sim$200 M31 clusters,  offering a great improvement 
in velocity accuracy over previous efforts.
This increase in accuracy lead a recent study by Morrison et al. (2004)
to kinematically identify what is possibly a {\it thin} disk 
population of M31 GCs. Such a system has not been 
observed in the Milky Way.

van den Bergh (1969) was the first to study the metallicity 
distributions of GCs in M31, from integrated spectra 
and $UBV$ photometry obtained with the Hale 200-inch. 
He concluded that on average, the M31 GC system was more
metal-rich than both the Milky Way and Fornax dwarf
GC systems.
In an important contribution, 
Burstein et al. (1984) comprehensively discussed
differences in the stellar content of GCs in the Milky Way and
M31. From integrated spectra, they found that at high
metallicity, at a  given metal-line strength the hydrogen lines 
(specifically H$\beta$) of the M31 GCs were systematically stronger 
than their Galactic counterparts. 
They explored a number of possible explanations for this
phenomenon, and concluded that the most likely explanation was
for an age difference in the GCs, in that the M31 GCs were
systematically younger than the Milky Way's.
They also found that CN in the intermediate- to metal-rich
M31 GCs was significantly enhanced (by several tenths
of a magnitude) when compared to elliptical galaxies. 
This CN enhancement could not be demonstrated at low
metallicities.
No satisfactory explanation which could explain both the 'CN' 
anomaly and the enhanced Balmer indices has been put forward.

In a series of papers Brodie \& Huchra 1990, 1991;  
Huchra, Brodie \& Kent 1991) investigated the
metallicity and abundances of M31 GCs using a 
more extensive sample of integrated spectra. 
In general, they found 
that the properties of the M31 GC system were 
similar to those of the Galactic system. Notable
exceptions were that the CN and H+K line-strengths were higher 
in the M31 GCs (at a given metallicity) than the Galactic
GCs, supporting the Burstein et al. (1984) findings.
 However, with the exception of a few individual cases, they
saw no systematic evidence for an enhancement in H$\beta$
in M31 GCs.

The CN enhancement in the M31 GCs is now reasonably 
well established (Trippico 1989; Ponder et al. 1998;  
Li \& Burstein 2004), whilst
the differences in H$\beta$ are somewhat more controversial (see 
Trippico 1994 for a short review). 
However, it is important to note that in the above studies, 
the most metal-rich Galactic GCs were not well represented, mostly due
to the difficulty of observing these objects against
the bright background presented by the Galaxy.
If metal-rich GCs are associated with spiral
bulges (Frenk \& White 1982; Cot$\acute{e}$ 1999; 
Forbes et al. 2001), then M31's bulge, which is an
order-of-magnitude more massive than that of the Milky Way 
should host many more metal-rich bulge GCs, resulting in an 
unequal comparison between the two GC populations (see
Trager 2004).

In this paper, we compare Galactic GC data, 
which include a number of 'bulge GCs' 
(Cohen, Blakeslee, \& Rhyzhov 1998; Puzia et al. 2002)
and newly analyzed spectroscopic data for M31 GCs in order to 
investigate some of these issues in detail. 
In a forthcoming paper (Beasley et al., in preparation (Paper II)), 
we compare these full datasets to contemporary stellar population
models in order to derive parameters such as metallicity, 
age and $\alpha$-to-iron abundance ratios.

The outline of this paper is as follows:
In Section~\ref{Data} we discuss the samples
under study and new data reduction steps 
required. Next, details of the spectroscopic
system are given in Section~\ref{Indices}.
In Section~\ref{Analysis}
we look at the behavior of the Balmer and CN indices
of the GCs, compared to recent elliptical
galaxy data. In Section~\ref{SecYoung}
we look in detail at eight young 
clusters in the M31 spectroscopic sample.
Finally, we present our discussion and conclusions
in Section~\ref{Discussion}. 

\section{Globular Cluster Data}
\label{Data}

\subsection{M31 Globular Clusters}
\label{M31Globs}

The integrated spectra of M31 globular clusters used in this study
were previously described in Barmby et al. (2000). These authors
obtained radial velocities for the clusters, and derived 
metallicities using the empirical calibrations of Brodie \& Huchra (1990). 
We have not used the MMT spectra described in Barmby et al. (2000), 
since these spectra are generally of insufficient signal-to-noise 
(S/N) for accurate line-strength analysis. 
In the following, we describe this dataset, and outline the
data reduction steps which differ from Barmby et al. (2000). 

Integrated spectra for 42 GC candidates in M31 
were obtained with the Low Resolution Imaging Spectrograph 
(LRIS; Oke et al. 1995) on the Keck-I telescope.
The observations were undertaken in 1995, September 21st-23rd, and
in 1996, December 8th, with some effort to obtain spectra
of clusters in the central galactic regions of M31.
Candidates were drawn from the catalog of Barmby et al. (2000), 
with the majority originating from the work of Battistini et al. (1987). 
Integrations were typically 5 minutes per target, 
which with characteristic magnitudes of V$\sim$17, yielded spectra
with S/N of 20 -- 200 per resolution element.
Typical seeing was 1 arcsec during these observations, and
the same 1 arcsec $\times$ 1.5 arcmin longslit was used for all
clusters. A 600 line mm$^{-1}$ grating was used yielding 
a 1.2~\AA\ pixel$^{-1}$ dispersion, and a full-width, half-maximum
resolution (FWHM) of 5~\AA. 
This resolution was determined from measuring a series
of isolated arc-lines in the comparison lamp spectra, and found to be
constant over our spectral range 3670--6200~\AA.

The data reduction steps of bias subtraction, flat-fielding and 
sky-subtraction were performed using standard packages in
$\textsc{IRAF}$\footnote{IRAF is distributed by the National 
Optical Astronomy Observatories, which are operated by the 
Association of Universities for Research in Astronomy, Inc., 
under co-operative agreement with the National Science Foundation}.
The spectra were wavelength calibrated using arc lamp spectra, covering
the full spectral range, which were interspersed amongst the M31 observations.

The accuracy of the relative flux calibrations of the spectra 
can have a significant impact on the measurement of certain broad 
line-strength indices. For these data, this is true of the Lick 
CN$_1$, CN$_2$, Mg$_1$, Mg$_2$ and TiO$_1$ indices.
For the present study, the original un-flux 
calibrated spectra from Barmby et al. have been re-fluxed 
using a larger series of flux standards taken during the 
time of the LRIS observations. These flux standards, BD+28$^\circ$4211, 
BD+25$^\circ$4655, Feige 110 and GD 50 were all selected
from the faint spectrophotometric standard star list of Oke (1990).
The standards were all taken with the same slit size 
(1 arcsec $\times$ 1.5 arcsec) as the program observations, 
and covered a range of airmasses.
In general, the new flux calibrations agree to within $\sim$10\% of 
the old calibrations. However, we do observe significant differences
in the spectra which exhibit earlier spectral-types (see Section~\ref{SecYoung}), 
which we attribute to differences in airmass between the original flux
standards and the program clusters.
The following analysis is based on the newly 
flux-calibrated M31 data.

Reliable line-strength measurements require a S/N  
such that the index and pseudo-continuum are well defined.
Therefore, we have focused on a high S/N ($>$30 pixel$^{-1}$) subset
of these data.
From 42 spectra of candidate M31 GCs, 36 are of sufficient S/N for
a detailed analysis of their line-strength indices. 
Five clusters, all drawn from the near-nucleus 
cluster catalog of Battistini et al. (1993),  
we have identified as being likely 
foreground G-dwarfs of the basis of their 
spectra. These objects exhibit weak CN and TiO 
molecular bands, closely following the dwarf
sequences of the Lick stellar library, rather
than that of the majority of the M31 GCs
and Lick giants. Moreover, 
a measure of their surface gravities based
upon the Sr{\small II}/Fe{\small I} diagnostic
(Rose 1985) places them clearly on the
locus of Lick dwarfs, rather than that of
a dwarf-giant mix expected for GCs 
(Tripicco 1989)\footnote{Although circumstantial, 
these clusters also possess a mean velocity of
--93 km$^{-1}$. This places them at the lower-velocity end
of the M31 GC velocity distribution, which
has a maximum-likelihood systemic velocity of
--284$\pm$9 kms$^{-1}$, and an overall velocity
dispersion of 156$\pm$6 kms$^{-1}$ (Perrett et al. 2002).}.
These objects (NB~68, NB~74, 
NB~81, NB~87 and NB~91) were removed from the sample.
The basic data for all the M31 clusters 
which satisfy our S/N and 'star cluster' 
criteria are listed in Table~\ref{BasicData}.
The spatial distribution of the sample with respect
to M31 is shown in Figure~\ref{Spatial}.

To anticipate some of the results of this study, 
spectra for eight of the M31 GC candidates have 
spectral characteristics typical of young stellar
populations or A-F stars; weak metal-lines, relatively 
strong absorption in the hydrogen Balmer lines 
and relatively blue continua when
compared to 'classical' G or K-type GC spectra.
These are discussed separately in Section~\ref{SecYoung}
and not included in the main analysis.

\subsection{Galactic Globular Clusters}
\label{GalacticGlobs}

Galactic GCs play a vital role in the interpretation 
of the integrated light of more distant GC systems. They 
not only allow for direct comparative studies, but also provide
important benchmarks for stellar population models since they
generally have independently derived ages, metallicities and
often, heavy-element abundance ratios. Increasingly, 
high-quality integrated spectra for Galactic GCs are
appearing in the literature, and we make use of 
two such studies.

Cohen et al. (1998; henceforth CBR98) 
obtained LRIS long-slit spectra for 
12 Galactic GCs in order to compare with their multi-slit M87 GC data.
These spectra were obtained by scanning $\sim$90 arcsec across the
cores of the Galactic GCs in order to synthesize a larger aperture, 
and thus obtain a representative integrated spectrum. 
These data are all of high S/N, but were taken using 
differing grating angles yielding 6 spectra with wavelength
ranges in the interval 3800 -- 6500~\AA, and 6 spectra slightly 
shifted to the red, with wavelength ranges of 4500 -- 7200~\AA. 
The LRIS 600 lines mm$^{-1}$ grating used yielded a dispersion
of 1.2~\AA\ pixel$^{-1}$, and resolution (FWHM) of $\sim$6~\AA.

CBR98 did not flux-calibrate these data, and measured indices
by individually shifting each index to its rest wavelength. 
We elected to derive new wavelength solutions for the 12 spectra (kindly
provided by J. Cohen and J Blakeslee, see Beasley et al. 2004)
by comparing the central index positions in these 
spectra with known wavelengths from our present M31 data, and 
Galactic GC data from Puzia et al. (2002). We have also performed a 
first-order flux calibration to the spectra 
using the sensitivity functions derived from our long-slit LRIS data
using a similar instrumental set-up\footnote{In the case of the CBR98
data, we found that the effect on the measured line-strengths using 
flux- versus un-flux-calibrated spectra was within the measurement 
uncertainties even for broad indices.}.

Puzia et al. (2002; henceforth P02) have presented high-quality
integrated spectra for 12 Galactic GCs, with the inclusion of a number 
of bulge clusters. Their sample also contains an integrated spectrum
of the outer bulge of the Galaxy.
These authors obtained fixed-position (i.e. not drift-scanned) 
spectra with different pointings (see P02) using the 
Boller \& Chivens Spectrograph at the ESO 1.52m at La Silla.
A 600 lines mm$^{-1}$ grating gave 1.89~\AA\ pixel$^{-1}$, 
with a fixed 3 arcsec slit yielding a resolution of 
$\sim$6.7~\AA. 

Four of the GCs in the P02 sample are in
common with those in the CBR98 sample, namely 
NGC~6356, NGC~6528, NGC~6553 and NGC~6624, resulting in a total of 20
unique GCs for our analysis. These four common clusters 
between the CBR98 and P02 datasets have allowed us 
to examine any systematic offsets between these data, in 
addition to estimating the true uncertainties in the 
CBR98 data (Section~\ref{Indices}).

\vbox{
\begin{center}
\leavevmode
\hbox{%
\epsfxsize=9cm
\epsffile{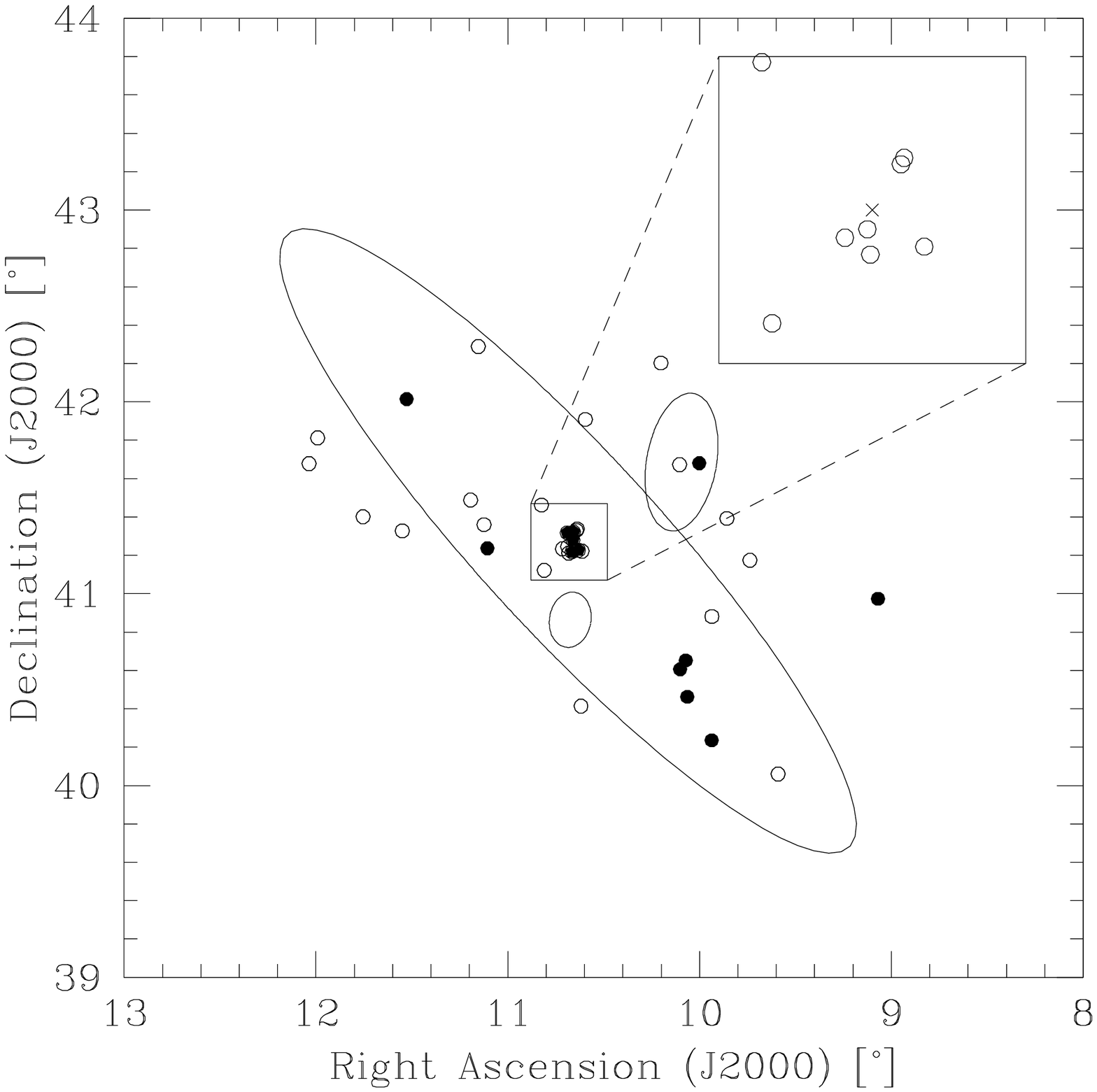}}
\figcaption{\small
Spatial distribution of sample globular clusters in M31. 
Open circles represent 'normal' globular clusters, filled
circles represent clusters which we identify as young
($<$ few Gyr). 
Ellipses represent faint isophotes of M31, M32 (lower left) and 
NGC~205 (upper right). Most GCs have a projected location 
in the disk/halo of M31, however several are concentrated 
in the galactic central regions, shown in the enlargement in the
upper right corner.
\label{Spatial}}
\end{center}}

\section{Line-strength Indices}
\label{Indices}

\subsection{Lick Indices}
\label{LickIndices}

We have chosen to investigate the chemical properties
of the M31 and Galactic globular clusters using the 
line-strength indices defined by the Lick group 
(Burstein et al. 1984; Trager et al. 1998).
For the following discussion, we first draw 
a clear distinction between the Lick ${\it indices}$ 
and the Lick ${\it system}$.

Lick {\it indices} now describe a set of 25 indices 
(see Trager et al. 1998 and Worthey \& Ottaviani 1997 for
the latest definitions) which are defined by two central
feature bandpasses, and four pseudo-continuum bandpasses
(a total of six bandpasses, or wavelengths).
However, the Lick {\it system} refers to Lick indices which are
measured on spectra which match (or mimic)
the behavior of the mildly non-linear and un-flux calibrated 
Lick/Image Dissector Scanner (Lick/IDS). This is achieved 
by virtue of either the spectra having been acquired with the Lick/IDS, 
or by calibrating onto the Lick/IDS system by matching its resolution
and correcting for systematic offsets through observing
sufficient Lick standard stars.

The advantages in using the Lick system and indices
are that they are widely used by galactic and globular cluster 
researchers, and that an extensive stellar library
exists for the construction of stellar population
models (e.g., Worthey et al. 1994). Important drawbacks
are that calibrating non-Lick/IDS data onto the Lick 
system can be difficult, generally requiring many
observations of standard stars, and that data of higher
resolution than that of the Lick library (8--11~\AA), must be
degraded to this resolution, potentially losing important
information.

In principle, line-strength indices measured on 
flux-calibrated spectra obtained with modern 
spectrographs and linear detectors
should be well reproducible, providing these data are at 
the same resolution (e.g., Vazdekis 1999). 
For the case of spectra 
with good sky-subtraction and flat-fielding, no
systematic offsets should be observed between measurements
of common objects, with the scatter entirely reflecting
the random uncertainties between the two observations.
Therefore, we have chosen not to correct these data 
to the Lick system, but compare the CBR98, M31 and 
P02 data at the Lick resolution.
Note for the P02 data, we have 'de-Licked' these
indices using the values given in P02. 
However, for comparison
with stellar population models based upon the Worthey et al. (1994)
Lick fitting functions, correction to the Lick system is a requirement.
We determine these corrections in Section~\ref{LickCorrections}.

To match the Lick/IDS resolution, the CBR98 and M31
data have been smoothed with a wavelength-dependent 
Gaussian kernel based on the 
known variation of FWHM with wavelength of the Lick/IDS given
in Worthey \& Ottaviani (1997). These M31 data were then shifted 
to the rest-frame using the velocities given in Barmby et al. (2000).
This shift was unnecessary for the CBR98 data since they were
wavelength calibrated in the rest-frame (Section~\ref{GalacticGlobs}).

%%%%%
 
CBR98 originally measured a subset of Lick indices
from their spectra using the original index definitions of
Burstein et al. (1984). For these re-calibrated CBR98 data, 
we have measured Lick indices using the updated definitions 
in Trager et al. (1998) and Worthey \& Ottaviani (1998).
Note that the differing grating angles used in these observations 
(Section~\ref{GalacticGlobs}) yielded different spectral ranges, 
resulting in 6 clusters which do not have measured indices 
blueward of 4500~\AA\ (C$_2$ 4668). All the measured
Lick indices which fall within the spectral range 
are given in Table~\ref{CBR98Indices}.

Due to the high S/N nature of these spectra, 
photon statistical errors generally underestimate the 
real uncertainty in the Lick indices measured.
Since we have no repeat observations of these clusters with which 
to determine our repeatability, we have estimated the 
uncertainty in each index by comparing our measured Lick indices
with four GCs which are in common with the P02 sample. 
We determine the rms difference between each 
Lick index to yield $\sigma_{\rm rms}$\footnote{In principle, this actually
provides an overestimation since we also include the intrinsic 
uncertainties within the P02 data. However, the
quadrature subtraction of these uncertainties in some cases leads
to an uncertainty of zero.}. The mean difference and $\sigma_{\rm rms}$
for each index between the common clusters are listed in 
Table~\ref{GalacticOffsets}. Unfortunately, only one of the Galactic
clusters common between the CBR98 and P02 datasets (NGC~6356)
includes all the blue indices in the CBR98 spectrum. Therefore, 
$\sigma_{\rm rms}$ for indices H$\delta_{\rm F}$ through to Fe4531
was taken to be the absolute difference between the measurement.
The shot-noise uncertainties for these data were determined using the 
method described by Cardiel (1998), which we 
denote $\sigma_{\rm S}$, yielding
a total error : $\sigma_{\rm T}^2 = \sigma_{\rm rms}^2 + \sigma_{\rm S}^2$.
Values of $\sigma_{\rm T}$ are also given in alternate rows
in Table~\ref{CBR98Indices}.

As can be seen, in general the agreement between the two datasets
is good and offsets between the CBR98 and P02 data are small.
This is true for the most important indices (e.g., CN, Mg$_2$, Mg $b$, 
Fe5270, Fe5335). We see significant offsets in G4300, H$\gamma_{\rm A}$, 
C$_2$4668 and Fe5015. The reason for these differences is unclear, 
however we note that the Lick correction employed by
P02 for these indices (which we have de-corrected these indices by)
are significant.

For the M31 data, twenty-four Lick indices were measured from
H$\delta_{\rm A}$ (the bluest) to TiO$_1$ (the reddest) as
defined in Trager et al. (1998) and Worthey \& Ottaviani (1998).
We were unable to measure TiO$_2$ since this index extends beyond 
our red wavelength limit. These indices are tabulated
in Table~\ref{M31Indices}.
We have estimated the uncertainties in these
data by comparing clusters in common between the two observing runs.
There are three clusters in common which are commonly used as 
velocity standards: 158-213, 163-217 and 225-280 (using the 
nomenclature of Huchra, Brodie \& Kent 1991), and we have taken 
the rms, $\sigma_{\rm rms}$, between index measurements of these clusters 
as an estimate of our repeatability. 

This method worked well for most indices, and any systematic
offsets were generally small. The exceptions to this were H$\delta_{\rm A}$
and NaD which both showed rather large offsets between the 1995, December
 and 1996, September runs (--0.4~\AA\ and --0.5~\AA\ respectively) for reasons 
that remain  unclear. In view of this we assigned rms's of 
0.4~\AA\ and 0.5~\AA\ respectively, and note 
that these are problematic indices. 
Poisson (shot-noise) errors ($\sigma_{\rm S}$) were calculated 
as for the CBR98 data.
Values of $\sigma_{\rm T}$ are given in alternate rows
in Table~\ref{M31Indices}.

\subsection{Corrections onto the Lick/IDS System}
\label{LickCorrections}

The usual method for correcting to the Lick/IDS system is to observe 
a range of standard stars in the Lick library which cover the metallicities
and range of spectral types of the stellar populations under study 
(Gonz$\acute{a}$lez, 1993). Ideally, these stellar spectra should be obtained
at the same time as the program observations, and crucially, 
obtained using the same instrumental setup.
Comparison with the Lick/IDS tabulated
values then provides the required corrections (linear or otherwise)
in order to mimic the Lick/IDS system.

Unfortunately, no such observations were taken for the CBR98 data, nor
for the M31 data. However, there are six Galactic GCs in
common between the CBR98 data and Trager et al. (1998), 
as are three M31 GCs in common between the M31 and Trager at al. 
(1998) data. The six clusters in common between the CBR98 data and
Trager et al. (1998) are NGC~6171, NGC~6205 (M13), NGC~6341 (M92), 
NGC~6356, NGC~6624 and NGC~6838 (M71). 
Trager et al. (1998) did not measure the higher-order Balmer
lines (H$\delta$, H$\gamma$) for these data, and therefore we have 
taken the indices measured on the Lick/IDS spectra from 
Kuntschner et al.(2002). Note that the Galactic and M31 cluster indices
given in Trager et al. (1998) were first published in Burstein et al. (1984), 
with the older Lick indices given in this paper.
In order to be consistent with the index bandpass definitions
in this study, all our comparisons are performed against
the newer measurements given in Trager et al. (1998).

To characterize the
relationships between the Lick and CBR98 index measurements, we have used
a simple least-squares linear fit; $I_{\rm Lick} = a + b I_{\rm CBR98}$, 
where $I_{\rm Lick}$,$I_{\rm CBR98}$ is index in question measured 
on the Lick and CBR98 data, and the coefficients $a,b$ are given in
Table~\ref{CBR98Offsets} along with the rms of the least-squares
fit. We have also calculated the mean difference
between these observations (Lick--CBR98) and their respective standard
deviations, which provide first-order corrections (i.e., additive corrections) 
and tabulate them in Table~\ref{CBR98Offsets}.
For several of the indices (G4300, Fe4383, Fe4531, C$_2$4668 and 
Fe5709) NGC~6171 showed large deviations from the 
mean, and these were removed from our calculations. We were unable to 
obtain any corrections for the bluest Lick index H$\delta_{\rm A}$.

Comparison between the Trager et al. (1998) and
CBR98 data shows significant deviations from unit slope for 
some indices (Table~\ref{CBR98Offsets}). 
In some cases, this may be attributable to resolution differences 
between the spectra (e.g., P02). 
We tested this by broadening the CBR98
data with wavelength-dependent Gaussian kernels several Angstroms
(FWHM) higher and lower than that of the Lick/IDS resolution, 
and re-measuring Lick indices. In general, we found that we could induce 
changes in slope of no more than 10\% in these indices with
the exceptions of Ca4227 and Fe5335, which are know to be very sensitive
to resolution (Vazdekis 1999).

For the M31 data, there are three clusters which are in common 
with the Trager et al. (1998) sample. This is not sufficient
to securely characterize any changes in slope. However these three 
clusters (158-213, 163-217, 255-280) do populate a reasonable range over
the metallicity distribution of M31 clusters (Brodie, Huchra \& Kent
give metallicities for these clusters of [Fe/H]\footnote{Unless
stated otherwise, [Fe/H] refers to metallicities on the 
Zinn \& West (1984) scale.} = --1.08$\pm$0.05, 
--0.36$\pm$0.27 and --0.70$\pm$0.12 respectively). Moreover, 
these three clusters are amongst the highest S/N spectra in
our M31 sample. In general the agreement between the M31 and 
Trager et al. (1998) indices is encouraging. We list the 
linear fit coefficients $a,b$, rms of the fit, mean values of 
(Lick--M31) and the standard deviations for each 
index in Table~\ref{M31Offsets}. 

\section{Analysis of the Main Sample}
\label{Analysis}

For the remainder of this paper, we use the
indices given in  Tables~\ref{CBR98Indices} and~\ref{M31Indices}, 
uncorrected to the Lick system.
Where there are clusters common between the CBR98 and P02 
Galactic data, we have used the P02 data due to their
more reliable error estimates.

\subsection{Balmer Lines}
\label{BalmerIndices}

Spinrad \& Schwiezer (1972), Rabin (1981) and later 
Burstein et al. (1984) found  evidence that
that M31 GCs were 
enhanced in the Balmer lines with respect to their Galactic
counterparts\footnote{Note
that although Rabin (1980) and Burstein et al. (1984) reached
similar conclusions regarding Balmer-line enhancements, 
in detail their respective H$\beta$ measurements show very poor agreement.}. 
Specifically, Burstein et al. (1984) found that at a 
given Mg$_2$ linestrength, the H$\beta$ index was on average 
higher than that of the Galactic globular clusters at 
Mg$_2$ $>$0.1.
These authors explored three possible origins 
for these differences--contributions from hot horizontal branch
stars, blue stragglers or younger ages--and favored this latter
interpretation.

Subsequent work by Brodie \& Huchra (1990, 1991) and Huchra et al. 
(1991) with a more extensive sample 
indicated that only a subset of M31 clusters appeared to have
such enhanced H$\beta$, and indeed this question
still remains controversial. 
We now examine this issue with the Galactic and M31 GC data.

\vbox{
\begin{center}
\leavevmode
\hbox{%
\epsfxsize=13cm
\epsffile{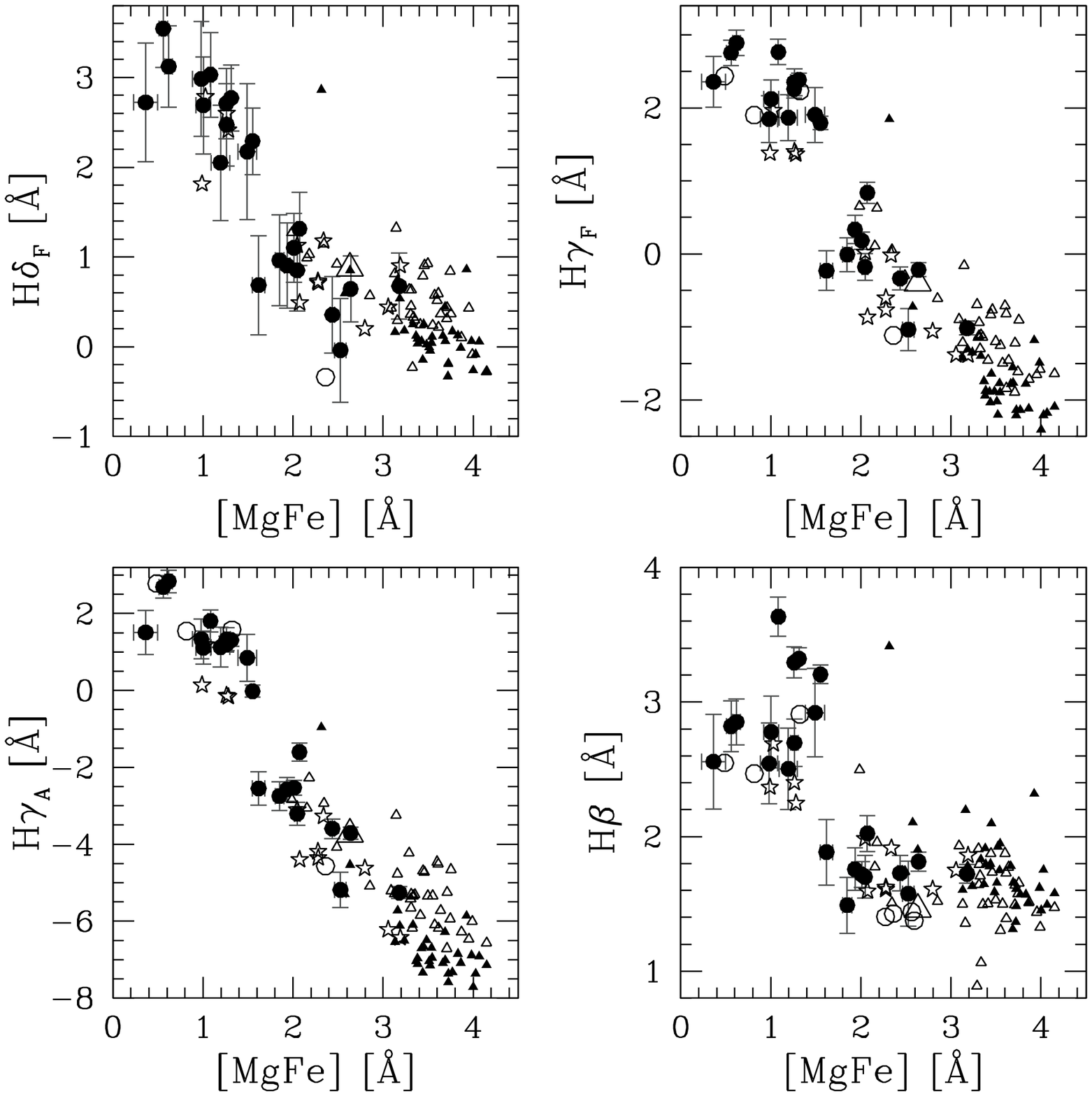}}
\figcaption{\small
Comparison between the M31 data (filled circles) and Galactic
GC data (open circles:CBR98; stars:P02) for the H$\beta$, 
H$\gamma$ and H$\delta$ Lick indices at a given [MgFe].
Uncertainties (grayed error bars) are only shown
for M31 data for the sake of clarity.
The open triangle indicates the outer Galactic 
bulge observations of P02.
Small open  and filled triangles represent 
data for Coma and field elliptical galaxies 
(S$\acute{a}$nchez-Bl$\acute{a}$zquez 2003).
For the metal-rich M31 GCs, we see no significant
enhancement in the Balmer lines compared to the Galactic data.
\label{Balmer}}
\end{center}}

In Figure~\ref{Balmer} we plot the Lick Balmer
indices of CBR98 and P02 compared to the M31 data at
given value of [MgFe]\footnote{[Mg$b$$\times$(Fe5270+Fe5335)]$^{1/2}$, 
Gonz$\acute{a}$lez (1993)}. 
We use [MgFe] since it 
is metallicity sensitive, relatively
insensitive to abundance ratio variations
(Gonz$\acute{a}$lez 1993; Thomas, Maraston \& Bender 2003), 
and the narrow indices Mg~$b$, Fe5270, and Fe5335
are not particularly susceptible to systematic
offsets between data.

The first thing to note in the four panels 
of Figure~\ref{Balmer} are the discontinuities in 
the plots at [MgFe] $\sim$1.6. These
gaps in the index-index relations reflect the well-established
bimodality of the Galactic and M31 cluster systems, a
gap which is not particularly evident in the 
Trager et al. (1998) data.
The presence of these discontinuities reflects the high S/N 
of these present data.
To contrast the Balmer indices of these data, we find it convenient
to separate the clusters into two groups, 
[MgFe] $<$ 1.5, and [MgFe] $\geq$ 1.5.
Using the empirical [MgFe]--[Fe/H] relation derived by P02, 
[MgFe]=1.5\AA\ corresponds to [Fe/H] $\sim$--0.8. 
We refer to these groups as the 'low-metallicity' and 
'high-metallicity' groups respectively.

Inspection of Figure~\ref{Balmer} shows no evidence
of systematic enhancements in H$\beta$, H$\gamma$ or 
H$\delta$ between the high-metallicity M31 and Galactic GC data.
Furthermore, there is no evidence
of a significant change in slope in any of the index-index
plots which was seen by Burstein et al. (1984).
Whilst there does appear to be a mild enhancement in several 
clusters at [MgFe]$\sim$2 , which is most
pronounced in their H$\gamma_{\rm A}$ and H$\gamma_{\rm F}$
indices, this is not particularly significant since there are 
two Galactic GCs (NGC~6388 and NGC~6637) at 
this metallicity which differ by $\sim$0.4~\AA\ in
H$\beta$; NGC~6388 lies squarely amongst these M31 
clusters.
The mean values of the H$\beta$ indices for all the
high metallicity M31 and Galactic clusters (P02 and CBR98 data) 
is 1.75~\AA\ $\pm$0.14~\AA\ and
1.63~\AA\ $\pm$0.20~\AA\ respectively. The straight mean, however, 
is not the fairest comparison in this case since the
slope of [MgFe]--H$\beta$ is non-zero, and there
are relatively more M31 clusters at [MgFe]$\sim$2
than there are Galactic clusters. The mean H$\beta$ of 
solely the metal-rich P02 data yields 1.74 $\pm$0.16~\AA\
which is identical to that of the metal-rich M31 GCs.

Turning our attention to the metal-poor clusters, 
the M31 GCs do appear to have somewhat
stronger H$\beta$ and H$\gamma_{\rm F}$ indices.
The mean H$\beta$ for these metal-poor M31
clusters is 2.95~\AA\ $\pm$0.35~\AA, whereas that
of the metal-poor Galactic clusters is 2.52~\AA\ $\pm$0.22~\AA.
This difference of 0.43~\AA\ in H$\beta$ is close to 
the dispersion in the metal-poor M31 clusters alone.
Such variations are not particularly unusual amongst metal-poor 
globular clusters, and possibly arise from differences in horizontal 
branch morphology (Lee et al. 2000; P02).

In Figure~\ref{Balmer} we have also plotted the high-quality
data for Coma and Field ellipticals from 
S$\acute{a}$nchez-Bl$\acute{a}$zquez (2003). 
Their sample consists of 59 galaxies from the 
field and Virgo cluster, and a further 34 galaxies from the central 
regions of the Coma cluster. These galaxies cover 
an absolute magnitude range $-22.5 < M_B < -16.5$, and 
comprise of dwarf to giant elliptical galaxies.
All the observations
were performed with an R$_e$/4 aperture and fully corrected to the
Lick system. We have 'de-corrected' these data (using 
the offsets kindly supplied by P. S$\acute{a}$nchez-Bl$\acute{a}$zquez)
in order directly compare with our data.

Interestingly, there appear to be obvious differences
between the loci of the field and Coma cluster galaxies
with respect to the GC data. 
The field galaxies generally appear to have lower
Balmer line-strengths when compared to the 
cluster galaxies at a given [MgFe].
Moreover, there is a suggestion that
the locus of the GC data more closely follows
that of the field galaxies than of those
in clusters. A straightforward (but non-unique)
interpretation
is that the GCs and field ellipticals
are generally older than cluster galaxies.
We note that the metallicities (as given
by [MgFe]) of the most metal-rich M31 and
Galactic GCs do not quite form a 
continuous relation with the galaxies.
However, as noted by P02, the metallicities
of the most metal-rich Galactic GCs are
comparable to the integrated Milky Way 
bulge light. This is consistent with the
idea that more metal-rich galaxies harbor
more metal-rich GC systems (van den Bergh 1975;
Brodie \& Huchra 1991; Strader, Brodie
\& Forbes 2004).

We see no difference in the Balmer-lines of the metal-rich
Galactic and M31 GCs at a given value of [MgFe], 
whereas previous studies (Burstein et al. 1984; 
Brodie \& Huchra 1991) have documented 
possible differences (at a given value of Mg$_2$).
What is the origin of this disagreement between studies?
One possible answer is S/N; these present data
represent an improvement in S/N over the older studies. 
For example, the mean uncertainty in our
H$\beta$ measurements are 0.17~\AA, whereas 
in the Burstein et al. (1984) study it is 0.26~\AA.
However, 
the effect of larger errors in e.g., H$\beta$ should 
in principle act to wash out
any systematic differences, rather than enhance them.
As shown in Table~\ref{GalacticOffsets} the agreement between 
the P02 and CBR98 data for the indices in question
is excellent (with the possible exception of H$\gamma_{\rm A}$, 
however in this case there is only one common cluster 
between these datasets). 
For example, the mean difference in the H$\beta$ index for
the four clusters in common is only (P02--CBR98) 
--0.0027, $\sigma$=0.13~\AA.

Indeed, in comparison with the CBR98 and P02 data, the 
H$\beta$ indices from Trager et al. (1998) for their
Galactic GC data look systematically low. This
is also evident when these data are compared to stellar
population models (Beasley et al. 2000; Kuntschner et al. 2003).
For the three clusters in common between the P02 data
and Trager et al.(1998), the difference is (P02--Trager)
0.44$\pm$0.28~\AA.
By comparison, the mean difference in H$\beta$ between the
M31 and Galactic clusters with MgFe $>$ 1.5 
in Trager et al. (1998) is $\sim$0.6~\AA, which lead
Burstein et al. (1984) to conclude that the M31 GCs 
may exhibit H$\beta$ enhancement.

It is hard to see how these differences have arisen, since
the H$\beta$ index is only weakly affected by continuum
slope changes or resolution differences.
Experiments involving adding synthetic noise to our
spectra, showed no signs of deviations from 
Gaussian errors.  
One possibility is foreground contamination from disk and/or
halo stars, however this is not expected to be a
significant effect (P02) and is unlikely to occur
for all the metal-rich objects in question.
Other possibilities include uncertain sky
subtraction especially in areas of high
background (CBR98, P02) and variations
in the sampled luminosity of the clusters (P02).
Based upon their thorough analysis of these issues, we
prefer the P02 indices over previous measurements.

In any event, on the basis of these present
data, we conclude that there is no Balmer-line enhancement in this 
sample of M31 globular clusters when compared to their 
Galactic counterparts
at high metallicity ([Fe/H]$\geq$--0.8). At low metallicity, 
there is some evidence for stronger H$\beta$, H$\gamma$ in the
M31 clusters which is perfectly consistent with 
horizontal branch variations.

\subsection{C and N in Globular Clusters}
\label{CNO}

One of the principle results of the Burstein et al. (1984)
study was the identification of strong CN molecular bands
($\lambda$4150~\AA) in the spectra of M31 clusters.
They found that at a given Mg$_2$ index strength, the CN feature
in the M31 clusters was up to 0.1 mag stronger than
that seen in elliptical galaxies. However, no
convincing comparison could be made with the Galactic GCs
due to the lack of the most metal-rich objects in their
sample.
A similar result was found by Brodie \& Huchra (1991)
using a similar definition of the near-UV cyanogen feature.

The lack of any significant variation in the G-band (CH) feature 
was indicative in both studies that C was not the main culprit 
in the CN-enhancement. Subsequent data, with the inclusion of the 
bluer NH feature at $\lambda$3360~\AA, (Tripicco 1989, 
Ponder et al. 1998; Li \& Burstein 2004)
strongly suggests the presence of enhanced N in the integrated spectra
of M31 clusters compared to Galactic dwarfs of comparable spectral type.

However, the Galactic globular cluster samples of Burstein et al. (1984), 
and Brodie \& Huchra (1991) did not include the most metal-rich Galactic GCs
in which one may  expect to see the strongest CN enhancements. Fortunately
the P02 data provide such a sample of clusters, and 
indeed these authors have noted that the CN strengths of the 
most metal-rich Galactic GCs
are significantly enhanced when compared 
to the integrated spectrum of the (outer) Galactic bulge.
Moreover, these authors also commented that these most-metal-rich 
Galactic GCs in their sample have similar CN strengths to 
the M31 sample of Trager et al. (1998).

\vbox{
\begin{center}
\leavevmode
\hbox{%
\epsfxsize=9cm
\epsffile{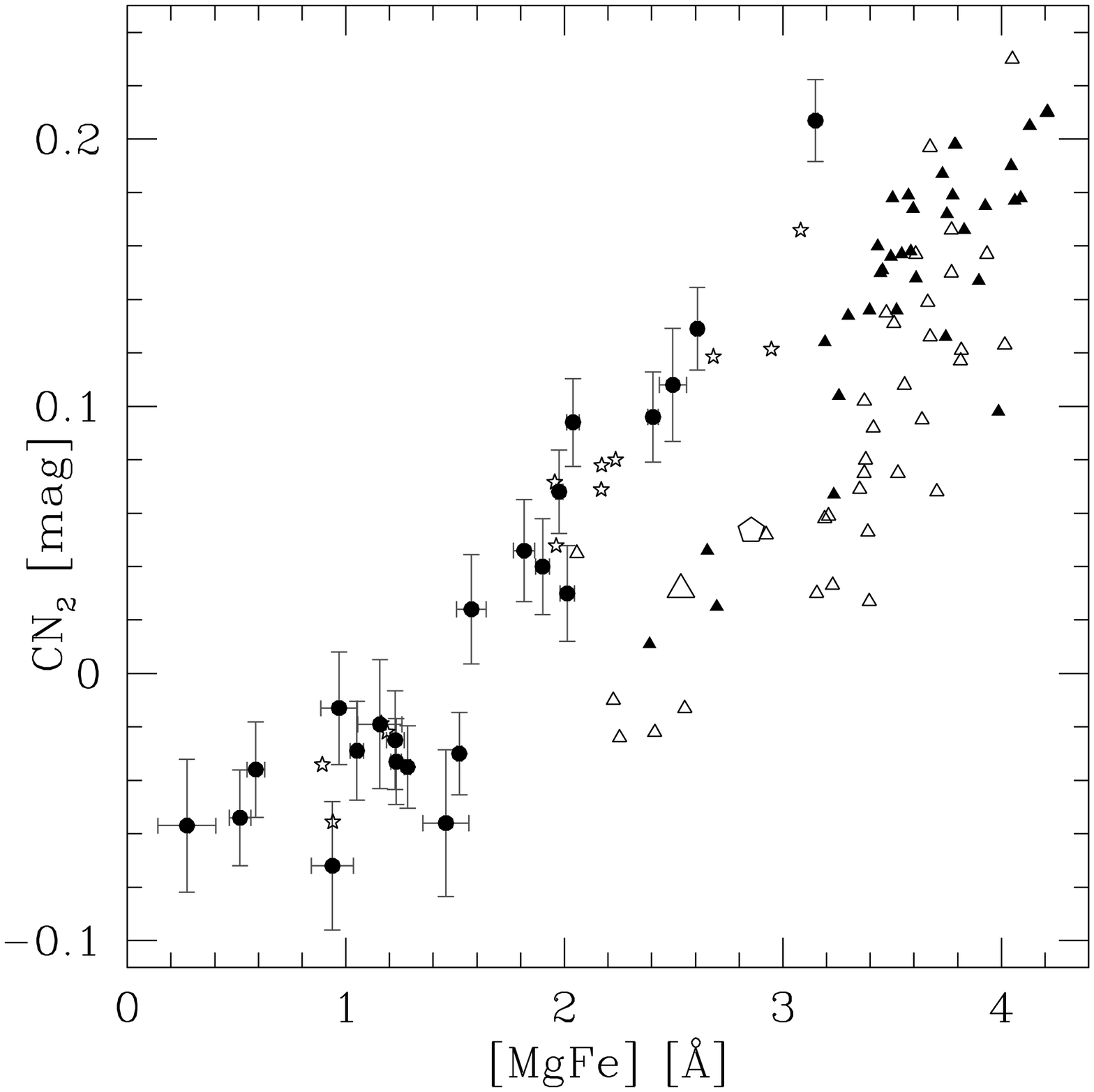}}
\figcaption{\small
The Lick CN$_2$ index as a function of [MgFe] for
the M31 GCs (solid circles with error bars), and P02 Galactic GC data 
(open stars). The open triangle
represents the mean Galactic bulge measurement from P02, and 
the pentagon denotes the M31 nuclear measurement of Trager et al. (1998).
Small filled and open triangles represent
data for field and Coma elliptical galaxies respectively 
(S$\acute{a}$nchez-Bl$\acute{a}$zquez et al. 2003).
\label{cn2}}
\end{center}}

In Figure~\ref{cn2} we show the Lick CN$_2$ index, for the M31
and Galactic GC data, at a given [MgFe].
We show the CN$_2$ index rather than CN$_1$, since
this latter index suffers from contamination
from the adjacent H$\delta$ feature.
Note that the CBR98 data generally 
do not reach down to $\sim$4000~\AA, and therefore we rely on the 
P02 dataset for comparison to these M31 data.
Figure~\ref{cn2} suggests that the CN indices for the 
Galactic and M31 clusters are
all significantly enhanced with respect to the 
bulges of the Milky Way and M31. This is 
consistent with the findings of Burstein et al. (1984) 
and P02. The difference in CN$_2$ between the 
Milky Way outer bulge observations and the M31 and P02
data at the same value of [MgFe] is $\sim$0.07 mag.

We have also plotted CN$_2$ data
from the field and cluster 
sample of ellipticals from S$\acute{a}$nchez-Bl$\acute{a}$zquez et al. (2003).
As discussed by these authors, 
the CN$_2$ indices of the field and Coma cluster ellipticals appear to show 
different behavior; field ellipticals appear
more enhanced in CN than those in the denser environment
of Coma. Indeed, the locus of the field galaxies is 
closer to that of the GCs, although there still appears
to be a substantial offset between the two.
S$\acute{a}$nchez-Bl$\acute{a}$zquez et al. (2003) concluded
that both N and C were enhanced in the field 
relative to cluster ellipticals, possibly through pollution
of the ISM through low- and intermediate-mass
stars.

We see no strong evidence for an
enhancement in CN$_2$ in the M31 clusters 
with respect to the Galactic GCs at any metallicity.
Similar behavior is seen in the CN$_1$ index. 
This is a rather surprising result, considering the fact
that the CN enhancement in M31 clusters has been 
well documented (Burstein et al. 1984; Tripicco 1989; 
Davidge 1990; Tripicco \& Bell 1990; Brodie \& Huchra 1991;
Ponder et al. 1998; Li \& Burstein 2003). 
Our findings do not appear to be an artifact of not
correcting to the Lick system, which actually exacerbates the problem 
since the correction is significant (0.032 mag for
CN$_2$). Through experimentation, we find that 
both CN$_1$ and CN$_2$ can
vary by up to several tenths of a magnitude depending
upon the details of our flux calibrations, 
but this is generally only true for the very metal-poor 
clusters which have relatively large amounts of blue
flux. This problem is particularly acute in the blue part of
the spectrum for broad indices, since the throughput
of the 'pre-blue-arm' LRIS roles off rapidly at $\sim$ 4000~\AA.

\vbox{
\begin{center}
\leavevmode
\hbox{%
\epsfxsize=9cm
\epsffile{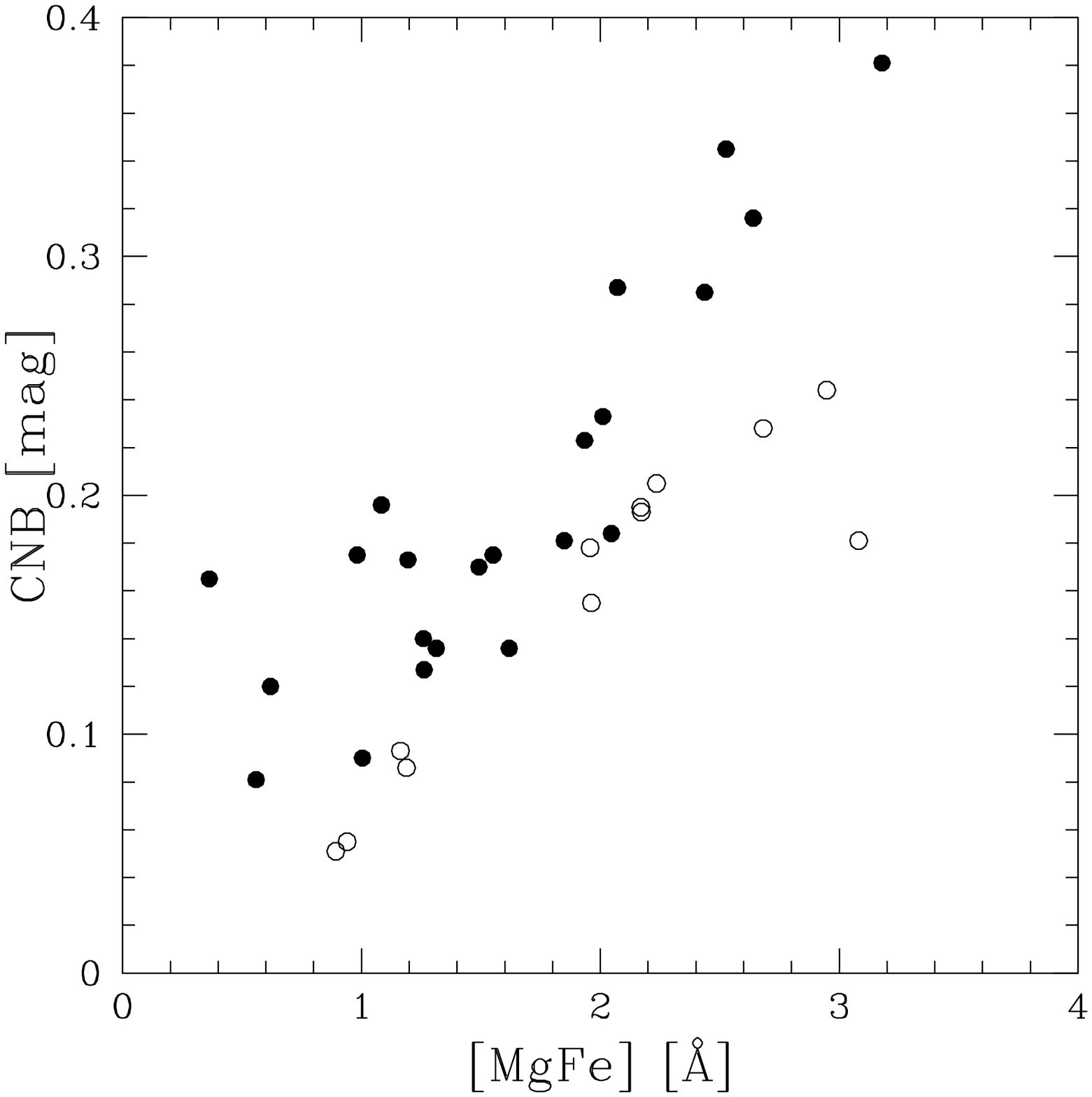}}
\figcaption{\small
The strength of the blue cyanogen index (CNB)
at a given [MgFe] for our M31 data (filled symbols)
and the Milky Way GC data of P02.
The M31 GCs appear clearly enhanced with respect to
Galactic GCs at all metallicities.
\label{CNB}}
\end{center}}

We have, however, measured the 
near-UV cyanogen index at $\lambda$3883~\AA\ 
(using the definitions given in Brodie \& Huchra 1990) 
for the M31 and P02 data, which through experimentation 
appears to be much less sensitive to our 
flux-calibration uncertainties\footnote{The 
truth of this statement is both data-dependent, 
and index-definition dependent. For example, the 
single-sided $\lambda$3883~\AA\ index often
used in stellar studies is prone to 
significant zero-point offsets 
(e.g.Cohen, Briley \&Stetson 2002).}.
The CN index at $\lambda$3883~\AA\ is also 
significantly stronger than the Lick CN indices.

The strength of CNB with [MgFe] is plotted
in Figure~\ref{CNB}. The difference in CNB between
the M31 and Galactic GCs is startling, and is
evident at all metallicities.
The metallicity range in which we can perform
a direct comparison between the M31 and Galactic
GCs is defined by the P02 data, with
--1.48 $\leq$ [Fe/H] $\leq$ --0.34.
At [Fe/H]$\sim$--1.5, the mean difference between 
the Galactic and M31 data is $\Delta$CNB $\sim$0.1 mag.
For the most metal-rich GCs at [Fe/H]$\sim$--0.4, 
 $\Delta$CNB$\sim$0.2 mag.

To check the reality of this result, since
Lick CN$_2$ and CNB appear to disagree, 
we have taken one of the 'CN-weak' 
Galactic GCs, NGC~6441, the spectrum of which was kindly 
supplied by T. Puzia, and divided it by one of 
the 'CN-strong' M31 clusters (225-280).
These clusters have very similar [MgFe]
strengths (i.e., very similar metallicities), 
and care was taken to ensure these
spectra were at the same normalization, resolution and spectral
sampling. This difference spectrum is shown in
Figure~\ref{ResidualCN}.

\vbox{
\begin{center}
\leavevmode
\hbox{%
\epsfxsize=9cm
\epsffile{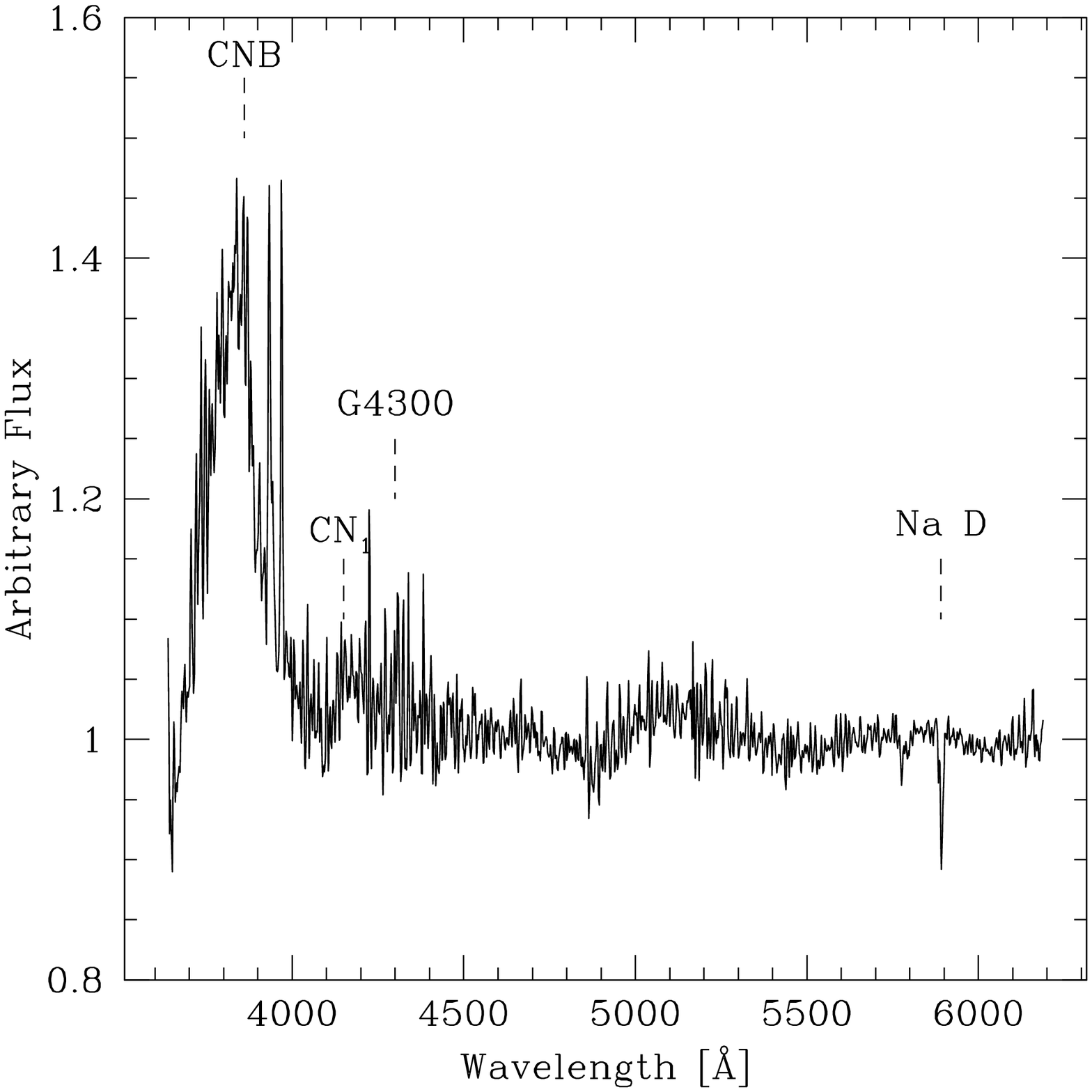}}
\figcaption{\small
Residual spectrum after the spectrum of NGC~6441, a 
Galactic GC, was divided by that of 225-280, 
a GC in M31. Both these clusters have [Fe/H]$\sim$--0.5.
The absorption system at CNB is clearly enhanced, indicating
greater CN absorption in the M31 GC. The Na D residual 
probably reflects the contribution of interstellar
sodium absorption in the spectrum of the Galactic GC.
\label{ResidualCN}}
\end{center}}

CNB in the residual spectrum is clearly enhanced, indicating
that the M31 GC has significantly stronger CN absorption
at the same metallicity than NGC~6441. Weak enhancement
is also seen in the Lick CN index, an enhancement which 
is washed out at lower metallicities. G4300 also 
appears to be mildly enhanced. Since this index
measures CH, this would suggest that 
C is also mildly enhanced (see below).

On the basis of the clear differences in the 
CNB index that we observe, we conclude that CN is 
enhanced in the M31 GCs with respect to
their Galactic counterparts at all metallicities.
The lack of obvious enhancement in the CN$_1$ and  
CN$_2$ indices between the Galactic and M31 
clusters probably reflects the relatively low
sensitivities of these indices to CN, when compared 
to CNB,  rather than any systematic offset between these
data (see Trager (2004) for a discussion of this issue).

Variations in CN are thought to be controlled
largely by variations in N, since free
C is rapidly incorporated into the CO
molecule.
Therefore, CN 'enhancement' is believed to reflect
an enhancement in N, this picture is supported
through observations of the NH molecule (Ponder et al. 1998; 
Li \& Burstein 2004).
However, can we rule out carbon variations?
Two Lick indices primarily sensitive to C are the 
C$_2$ swan band C$_2$4668 (formerly Fe4668) and 
G4300 which measures CH.

\vbox{
\begin{center}
\leavevmode
\hbox{%
\epsfxsize=9cm
\epsffile{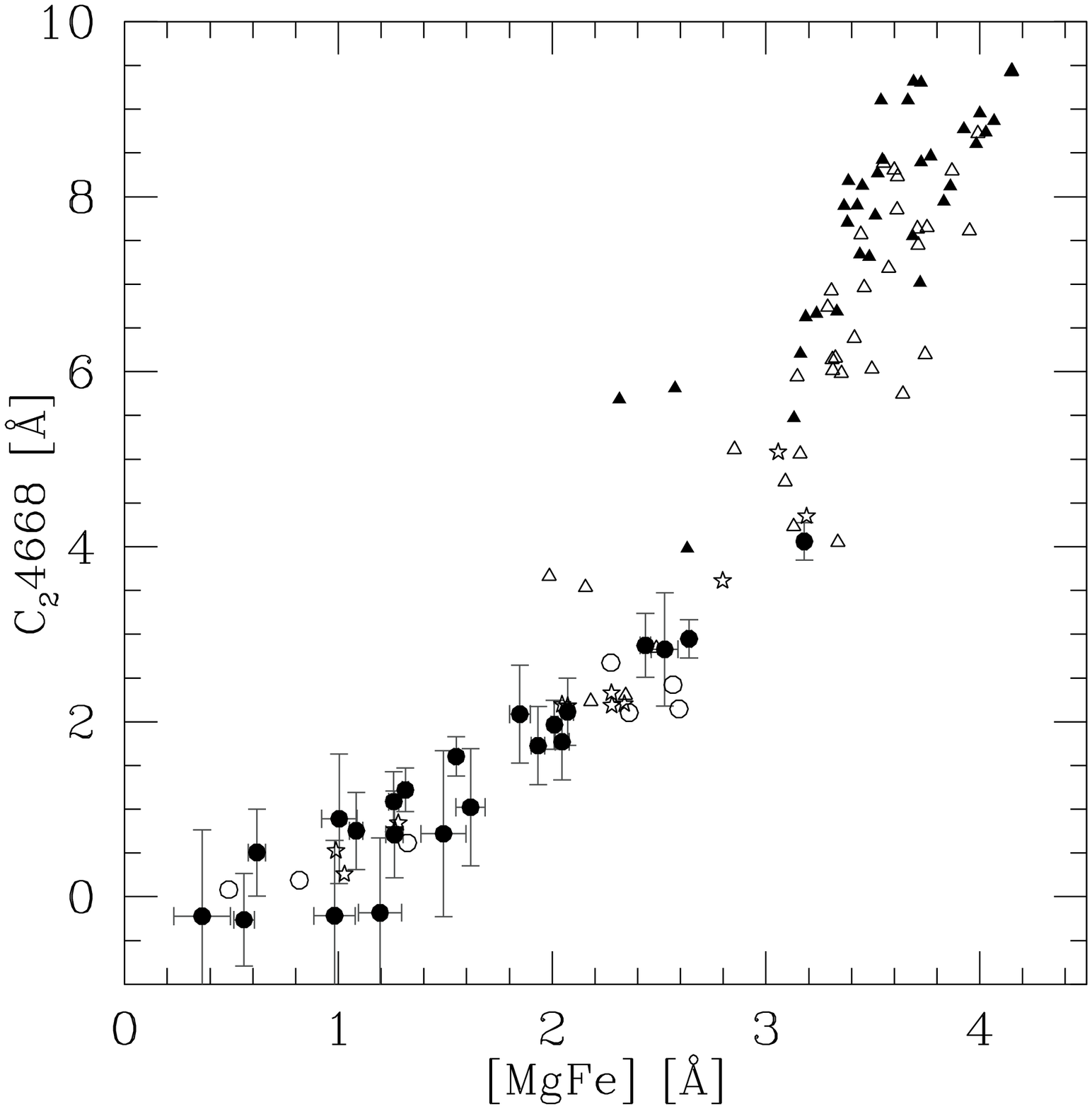}}
\figcaption{\small
The Lick C$_2$ 4668 index as a function of [MgFe].
Symbols are the same as for Figure~\ref{cn2}.
\label{C4668}}
\end{center}}

\vbox{
\begin{center}
\leavevmode
\hbox{%
\epsfxsize=9cm
\epsffile{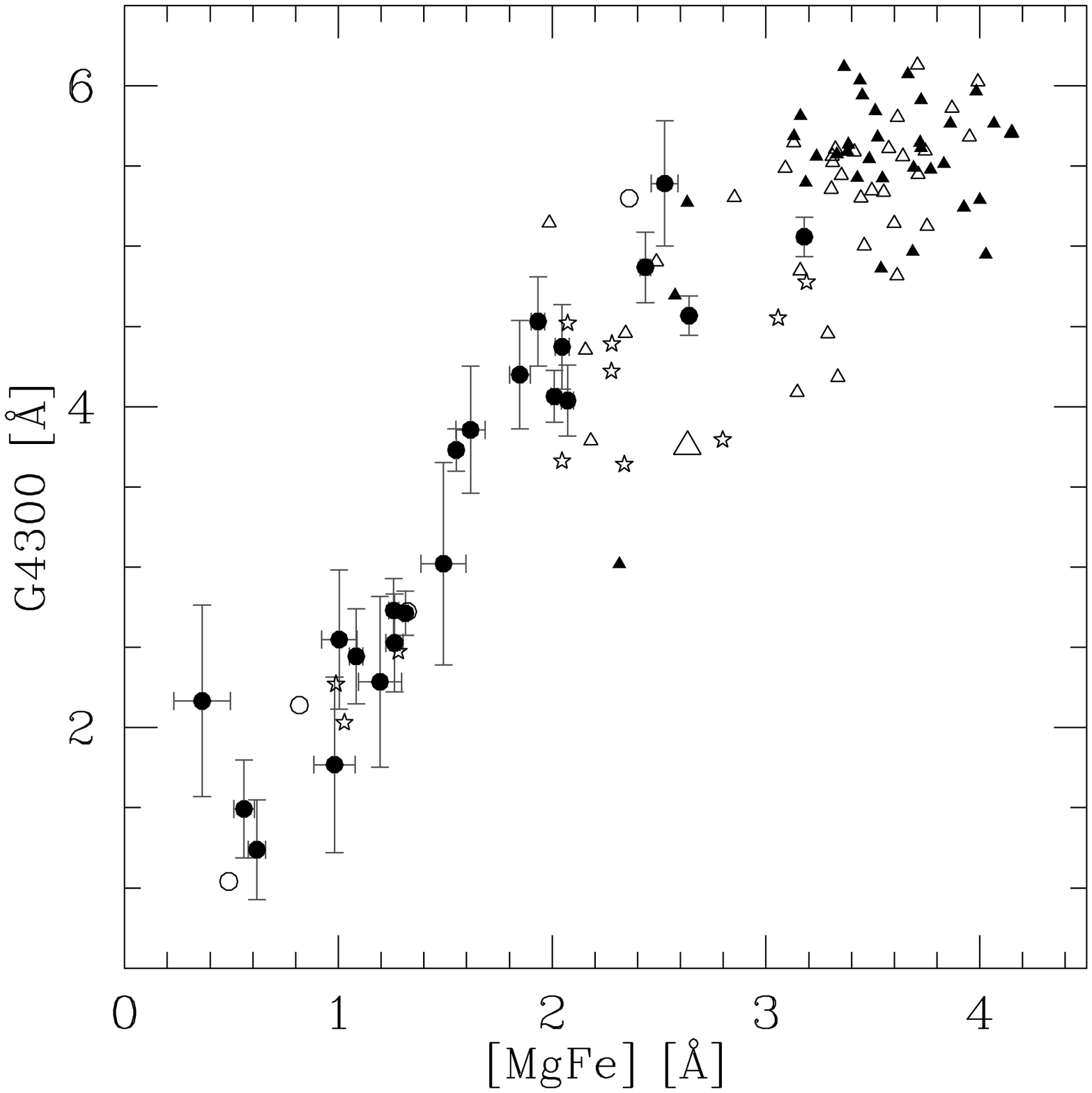}}
\figcaption{\small
The G4300 (CH) index as a function of [MgFe].
Symbols are the same as for the previous figure.
\label{CH}}
\end{center}}

In Figure~\ref{C4668} we plot C$_2$4668 as a function of [MgFe].
These data define a very tight sequence in this
plot, with a clear change of slope at [MgFe]$\sim$ 3, 
the transition from GCs to galaxies. On the basis
of Figure~\ref{C4668}, we see no evidence for
differences in C amongst the GCs. 
However, as discussed by 
S$\acute{a}$nchez-Bl$\acute{a}$zquez et al. (2003), 
this index is enhanced in field ellipticals with 
respect to cluster ellipticals at a given metallicity.
The behavior of the other carbon sensitive index, G4300, is 
plotted in Figure~\ref{CH}. 
In this case, there is
a suggestion that at [MgFe]$>$2, the M31 GCs 
do have a stronger G4300 indices than the 
Galactic GCs. The M31 clusters also have significantly
stronger G4300 than the outer bulge observations
of the Milky Way. Unfortunately, the interpretation
of such variations in G4300 are complicated by the fact
that this index is also sensitive to age and the 
O and Ti abundances.

\section{Young Star Clusters in M31}
\label{SecYoung}

Eight of the initial 42 GC candidates in the M31 sample 
exhibit mid-A to early F-type spectra, as would be expected of 
A-/F-stars or young star clusters. 
The existence of very young clusters in the M31 system
has been mentioned previously; Huchra, Brodie \& Kent (1991), 
and Barmby et al. (2000).

One of these clusters, 324-051, 
is also designated Hubble~V in NGC~205, and was taken 
as a reference template for the other candidate young
clusters in M31. Hubble~V was identified by 
Da Costa \& Mould (1988) as being a young object
from 3~\AA\ resolution optical integrated spectra.
Based on its similarity to the Small Magellanic Cloud
cluster NGC~419, Da Costa \& Mould (1988) concluded that 
Hubble~V was $\sim$1 Gyr old, but could not rule
out ages several Gyr older than this, or as young
as 100 Myr.

From the broad Balmer lines, weak metal-lines and relatively 
blue continua of the cluster spectra in our sample, 
we assign the clusters spectral types A5--F0. 
Such $\sim$2$\Msun$ stars which dominate the
main-sequence turn-off will complete 
core-hydrogen burning after approximately 1 Gyr 
(Schaller et al. 1992), consistent with the age of 
Hubble~V found by Da Costa \& Mould (1988)

The spectra of these cluster candidates are shown in 
Figure~\ref{YoungSpectra}.
Clear evolution in several indices can be seen 
as we proceed to later types in the figure; the Balmer lines
all weaken, the ratio of Ca II K/Ca II H+H$\epsilon$
increases as metal-lines strengthen and H$\epsilon$
weakens, and the G-band (G4300 in the Lick definition)
which measures the CH molecule starts to appear
in the Hubble V spectrum and later types.
The slope of the spectra also noticeably change, from bluer to
redder, although reddening also seems to be playing a
role here (e.g., 321-046). The clusters 222-227
displays a rather unusual shape, possibly due to heavy reddening, 
leading to its 'peculiar' designation in 
Table~\ref{BasicData}. It also happens to be our lowest 
S/N spectrum.

\vbox{
\begin{center}
\leavevmode
\hbox{%
\epsfxsize=12cm
\epsffile{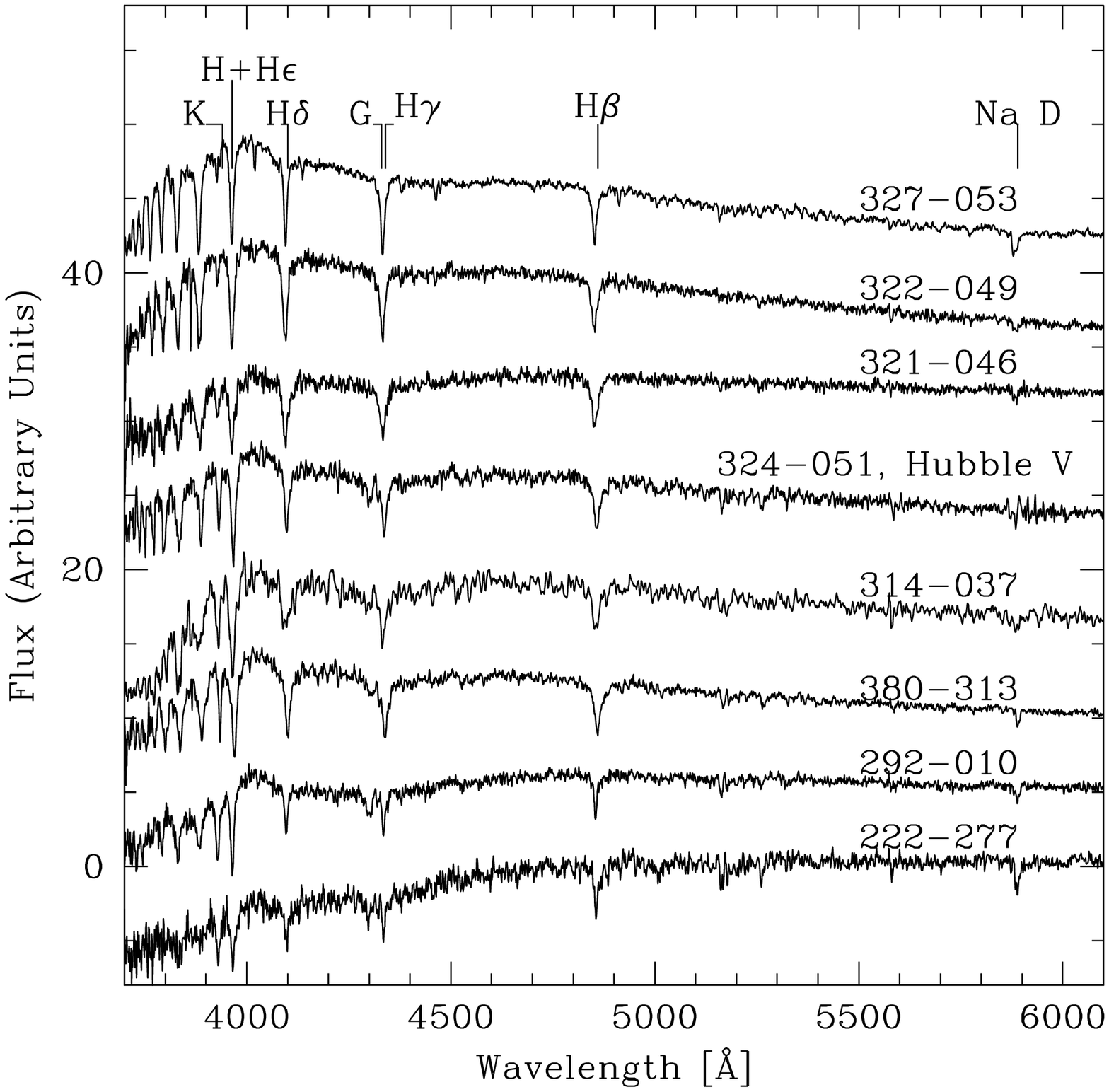}}
\figcaption{\small
LRIS spectra at $\sim$5.6~\AA\ resolution for the eight 
candidate young globular clusters in the 
M31 sample. The spectra have been normalized, with additive constants
on the $y$-axis for display purposes. The spectra are arranged
in approximate order of spectral type, running from earlier (A5)
to later types (F0; top to bottom).
\label{YoungSpectra}}
\end{center}}

The first obvious question to ask of these young cluster
candidates is : are these really globular clusters 
belonging to M31? 
In the original Bologna group catalogue
(Battistini et al. 1987), which was based on 
photographic plates, 
all the clusters for which we have spectra 
are designated A- or B-classes, i.e., high-confidence clusters, 
on the basis of their non-stellarity and circularly symmetric brightness
distributions. This effectively rules out 
that these objects are foreground
Galactic blue horizontal branch stars, or blue
stragglers.

As evidenced by the Large Magellanic Cloud (LMC)
system of star clusters, the separation of open and 
globular clusters is not always
clear (e.g., Hodge \& Wright 1967).
However, the distinction made is usually
one of mass and compactness.
The absence of high-resolution imaging 
for all the candidates precludes the secure
identification of these objects\footnote{We are currently
pursuing such imaging for these objects}.
However, HST/WFPC2 imaging does exist for Hubble V
(cycle 6, proposal ID 6699, PI D. Geisler).
We show the PC chip image of this object
in the F814W filter in Figure~9.
The cluster is clearly resolved into stars,
and Kim et al. (2002) have measured surface 
brightness profiles for this cluster 
finding that it is better fit by a King (1966)
model than a power a law which is 
usually applicable for open clusters 
(Elson, Fall \& Freeman 1987).
Kim et al.'s derived concentration parameter
of $c$=0.5 is comparable to the small
Milky Way cluster AM-4 (Madore \& Arp 1982).
The mass to light ratios 
of young clusters evolve rapidly due to the short main-sequence
lifetimes of massive stars, this in turn strongly effects
any mass estimates for the clusters.
We return to estimating the cluster masses later 
when we have more secure age determinations for these objects.

\vbox{
\begin{center}
\leavevmode
\hbox{%
\epsfxsize=6cm
\epsffile{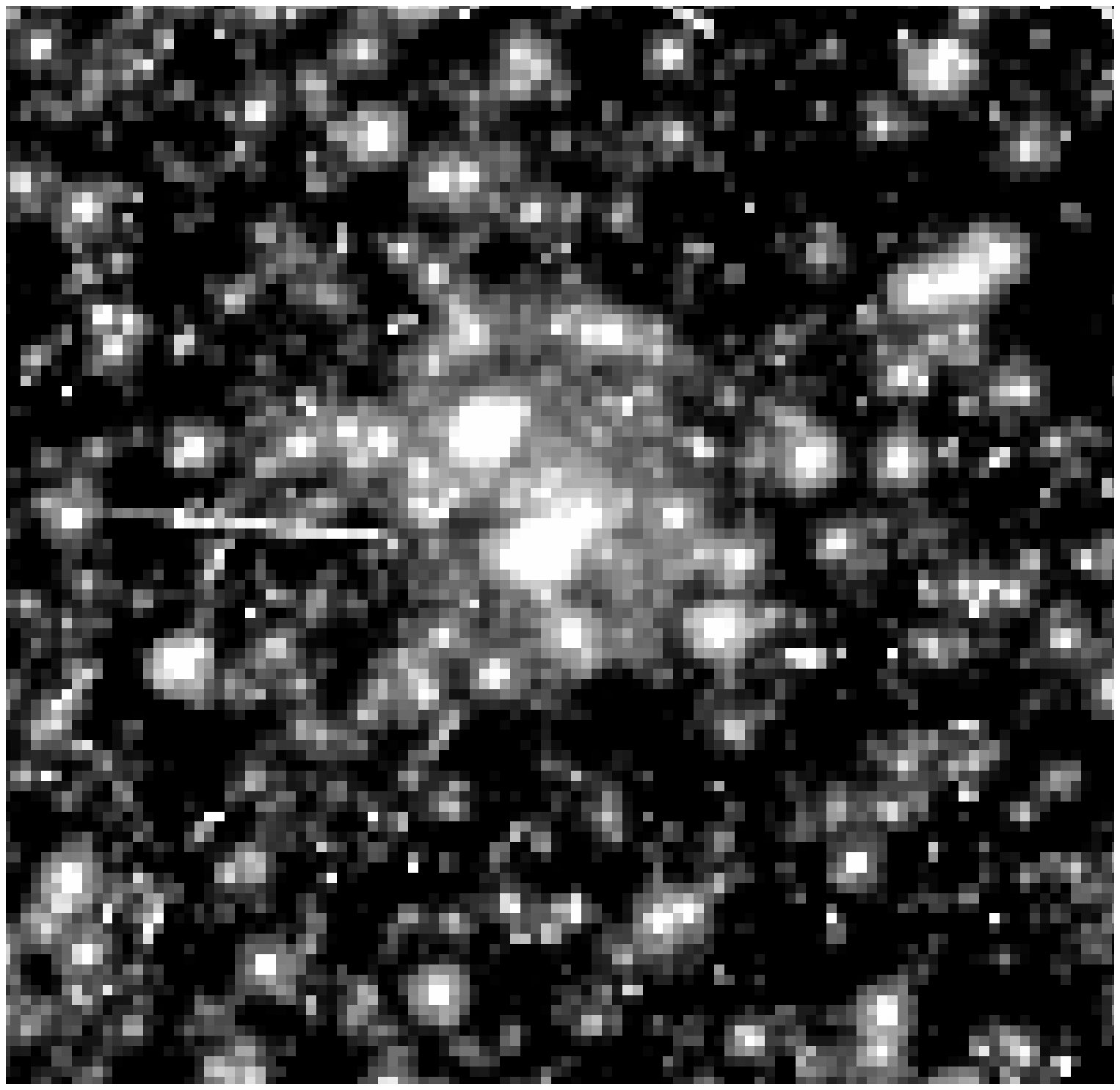}}
\figcaption{\small
HST/PC F814W image of the young cluster in NGC~205, Hubble V. 
The image is 5$\arcsec$ across, which, at the 
distance of M31, corresponds to a physical diameter of $\sim$17pc.
The pixel scale of the PC chip is 0.045 arcsec/pixel.}
\label{HubbleImage}
\end{center}}

Age-dating young stellar populations
($<$ 1 Gyr) from integrated spectra presents a different set of 
challenges from older stellar systems. For stellar populations older
than $\sim2\times10^8$ yr, Balmer lines decrease monotonically
with time and may be used to estimate the age of the 
system (e.g. Bruzual \& Charlot 2003). The key difficulty lies
in determining the metallicity of the stellar population, since
metal lines become very weak in young objects 
(e.g., Larsen \& Brodie 2002).
We have taken two approaches: first we compare the M31 spectra
with a series of LMC templates (Beasley et al. 2002) 
which bracket the expected age range, and then use 
stellar population models to provide
independent age and metallicity information on the clusters.
The two approaches are complementary, and provide useful insight
into the reliability of each method.

The LMC cluster system consists of clusters with a range
of metallicities and ages, from very metal-poor 'classical'
globular clusters to intermediate-aged and young, near-solar 
metallicity star clusters (e.g., Olsen et al. 1998; Geisler et al. 2003; 
Piatti et al. 2003)\footnote{The age-metallicity relation
for the LMC clusters in not continuous. It has been known for
some time that there is an 'age-gap' between 3$\sim$10 Gyr
(e.g., Jensen, Mould \& Reid 1988) with a corresponding 
'metallicity gap' at [Fe/H] $\sim$--1 (Olszewski et al. 1991).
This, however, is not a concern for the present study since 
we are interested in LMC cluster templates which are younger
than a few Gyr.}.
This range of cluster types provides an excellent source
of templates for more distant studies (Leonardi \& Rose 2003).
Beasley et al. (2002) published integrated spectra of 24 LMC 
star clusters with 'SWB-types' (Searle, Wilkinson \& Bagnuolo 1980; 
hereafter SWB) 
ranging from IVA--VII. Whilst the exact age-calibration 
of the SWB system is still a source of uncertainty, using the 
calibration of Bica, Claria \& Dottori (1992), these types
correspond to ages 0.20$\sim$16 Gyr.
The earlier calibration of Cohen (1982) gives somewhat
older ages for the intermediate SWB-types (III--VI).
 
Spectra for SWB I--III clusters (0.01--0.20 Gyr) were also 
obtained by Beasley et al. (2002),
but remained unpublished since the stellar population models 
assessed did not extend to such young ages.
We have used the full sample of Beasley et al. (2002) which 
consists of 65 clusters, the majority of which are less than 1 Gyr
old. In order to obtain the best match between the young M31 clusters
and the LMC templates, we cross-correlate each M31 spectrum
against the entire LMC library, and select the LMC 
cluster which yields the highest (normalized) cross-correlation
peak height. For this purpose, the M31 spectra were 
smoothed and re-sampled to match the natural resolution
of the LMC data (6.5~\AA). The spectra were then normalized by
a low order polynomial (since the LMC data are not flux calibrated), 
and cross-correlated in the wavelength interval 4000-5400~\AA, which
is the effective wavelength range of the LMC data.

To test the method, we first cross-correlated the M31 cluster
324-051 (Hubble V) against the LMC library, since we have 
some independent age information for this cluster. Da Costa
\& Mould (1988) determined an age of $\sim$1 Gyr for this cluster, 
with a lower bound of 100 Myr, whilst Lee (1996) determined
from his measured color of the cluster (B--V$_0$=0.33)
that the cluster is closer to 300 Myr.
The best correlation template we obtain for Hubble V is 
NGC~1940, an SWB II cluster with B--V$\sim$0.3 
(Bica et al. 1992). SWB II corresponds roughly to 
ages 30--70 Myr, which is surprisingly young. 

The normalized spectra of Hubble V and NGC~1940
are shown in Figure~10.
The two clusters look rather similar, and this is supported
by examination of the residual spectrum (dividing the Hubble V 
spectrum by that of NGC~1940). Apart from the mild 'P-Cygni'
profile near H$\gamma$, the residuals are relatively modest.
There also appears to be a slight enhancement in the CN
residuals, in the sense that CN is lower in Hubble V, although
the reality of this is unclear at present.

\vbox{
\begin{center}
\leavevmode
\hbox{%
\epsfxsize=9cm
\epsffile{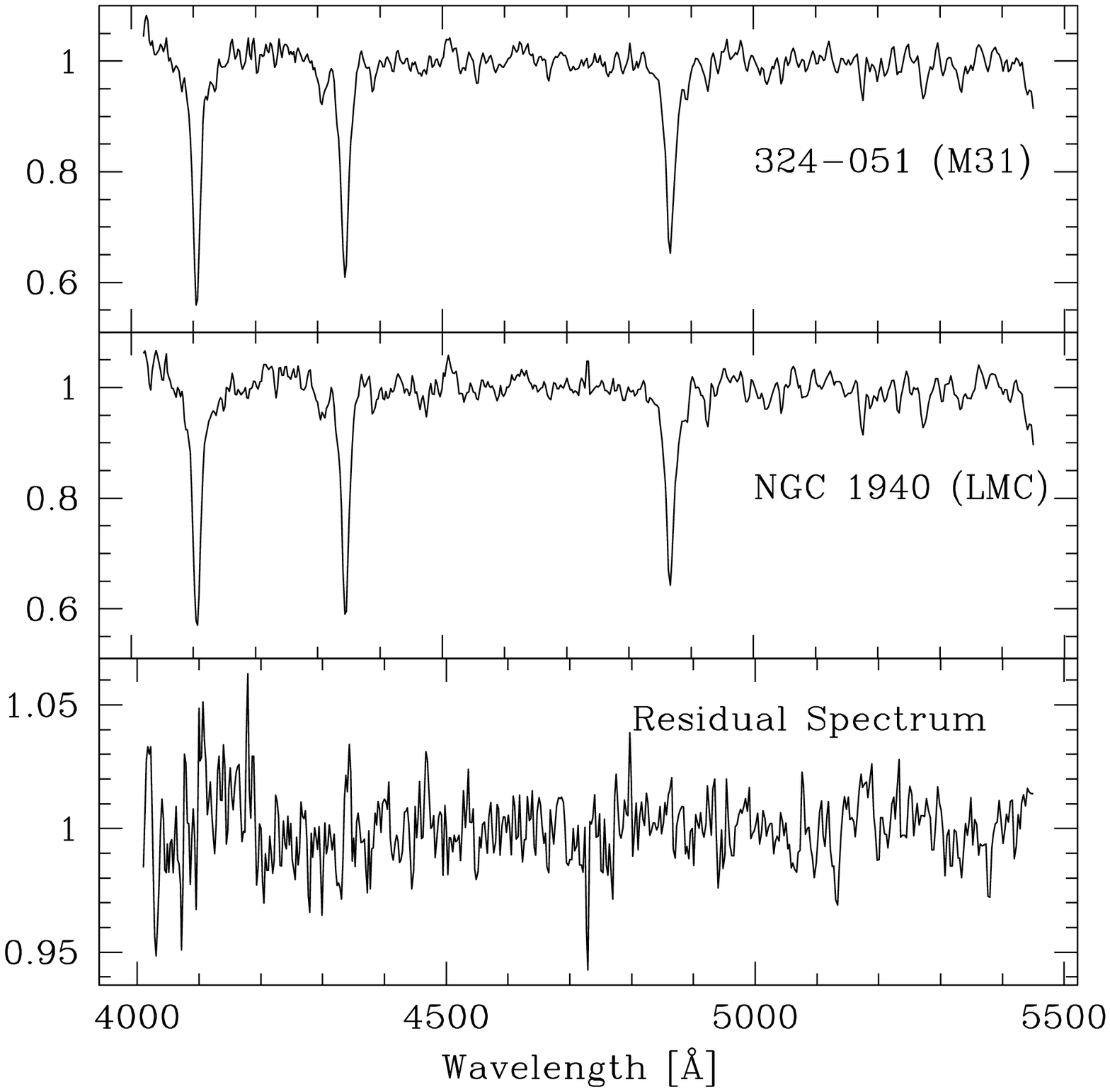}}
\figcaption{\small
Comparison of the normalised spectra of 324-051 (Hubble V; top panel) and
NGC~1940 (middle panel), which yielded the highest cross-correlation peak
of all the LMC library (Beasley et al. 2002).
The bottom panel shows the residual spectrum after
dividing 324-051 by NGC~1940.
}
\label{CrossCorr}
\end{center}}

The next closest templates are
NGC~2127 and NGC~2041 which are SWB III and II types
respectively, corresponding to an age range of 30--200 Myr
using the Bica et al. (1992) calibration.
Girardi et al. (1995) obtained an age of $\sim$100 Myr
for NGC~2041 based upon a calibration of the 
photometric $S$-parameter, slightly older
than the Bica et al. calibration predicts.
The work of Cohen (1982) suggests SWB III-types
are closer to 500 Myr. 
This is an appreciable age-range,
particularly when one considers that a 30 Myr cluster
will be dominated by B stars, whereas in a 500 Myr 
stellar population A/F stars will be the dominant source of
light. The spectrum of Hubble V 
(Figure~\ref{YoungSpectra}) suggests a population of
stars cooler than B, closer to late-A, due to the presence of 
metal-lines (particularly G-band) in the spectra, and the roll-over
in the spectrum at $\sim$4000~\AA. On the basis
of the spectrum alone, the cluster is unlikely to be significantly
younger than 100 Myr.
Moreover, it should be noted that if the LMC templates are more
metal-poor than the M31 clusters (a distinct 
possibility considering the mass-metallicity relation
of galaxies), then 
due to age--metallicity degeneracy a younger, more 
metal-poor LMC template will yield the best correlation match
with a slightly older, more metal-rich M31 cluster.

To conclude, we obtain an age-determination for Hubble V of 
100--500 Myr, in good agreement with previous determinations, 
under the assumptions that the chemical composition of the young M31
clusters (metallicity, abundance ratios) are similar to those
of the LMC system.

We have performed the previous exercise for the remaining 
young M31 spectra, each time choosing the three closest
cross-correlation templates to the cluster in question.
Our results are summarized in Table~\ref{CrossCorrTab}.
According to the LMC templates, the M31 clusters
span a significant range of ages from between 
10--30 Myr (327-053) up to 0.8--2 Gyr (292-010).
The relative age rankings are qualitatively supported by the 
appearance of the cluster spectra in Figure~\ref{YoungSpectra}.

We now appeal to the predictions of stellar population
models for more information regarding these objects.
To estimate the cluster ages and metallicities 
we have employed the evolutionary 
population synthesis models of 
Bruzual \& Charlot (2003; hereafter BC03).
In the optical range which we are interested in
(3670--6200~\AA), the BC03 models employ the 
recent stellar library (STELIB) of Le Borgne et al.(2003).
The library, which consists of 249 stars taken at
a resolution of 3~\AA\ (FWHM), covers with 
varying degrees of completeness the metallicity
range --2.0 $<$ [Fe/H] $<$ +0.50, luminosity classes
I to V, and spectral types O5 to M9.
Perhaps the most serious limitation of these (and other)
models for our purposes is the lack of hot stars 
($T_{eff}>$ 10,000 K) at non-solar metallicities.
This, however, should not be too problematic
since we expect the abundances of the young M31 
clusters to be at or near solar metallicity.

We show in Figure~11 the evolution of the Lick-defined
H$\beta$ index as a function of age and metallicity.
The models shown adopt the Chabrier (2003) IMF, with
a lower mass cut-off of 0.1$\Msun$ and upper
mass cut-off of 100$\Msun$. The use of a Salpeter
IMF has small effect on the optical indices in
question here (BC03).
According to the BC03 models, the H$\beta$ index increased rapidly
with time until $\sim$300 Myr, as turn-off A-stars come to dominate the 
integrated light. The strength of the index then decreases monotonically
with age. In this regime, age is the single-most 
important influence on this index, whereas metallicity
is a significant, but less important contributer.
We have plotted our H$\beta$ measurements for the eight 
young clusters in  Figure~11, which range from 
3.12 -- 6.83~\AA. In themselves, these suggest an age range
between 0.02--5 Gyr. The higher-order Balmer lines show
very similar behavior to that of H$\beta$.

\vbox{
\begin{center}
\leavevmode
\hbox{%
\epsfxsize=9cm
\epsffile{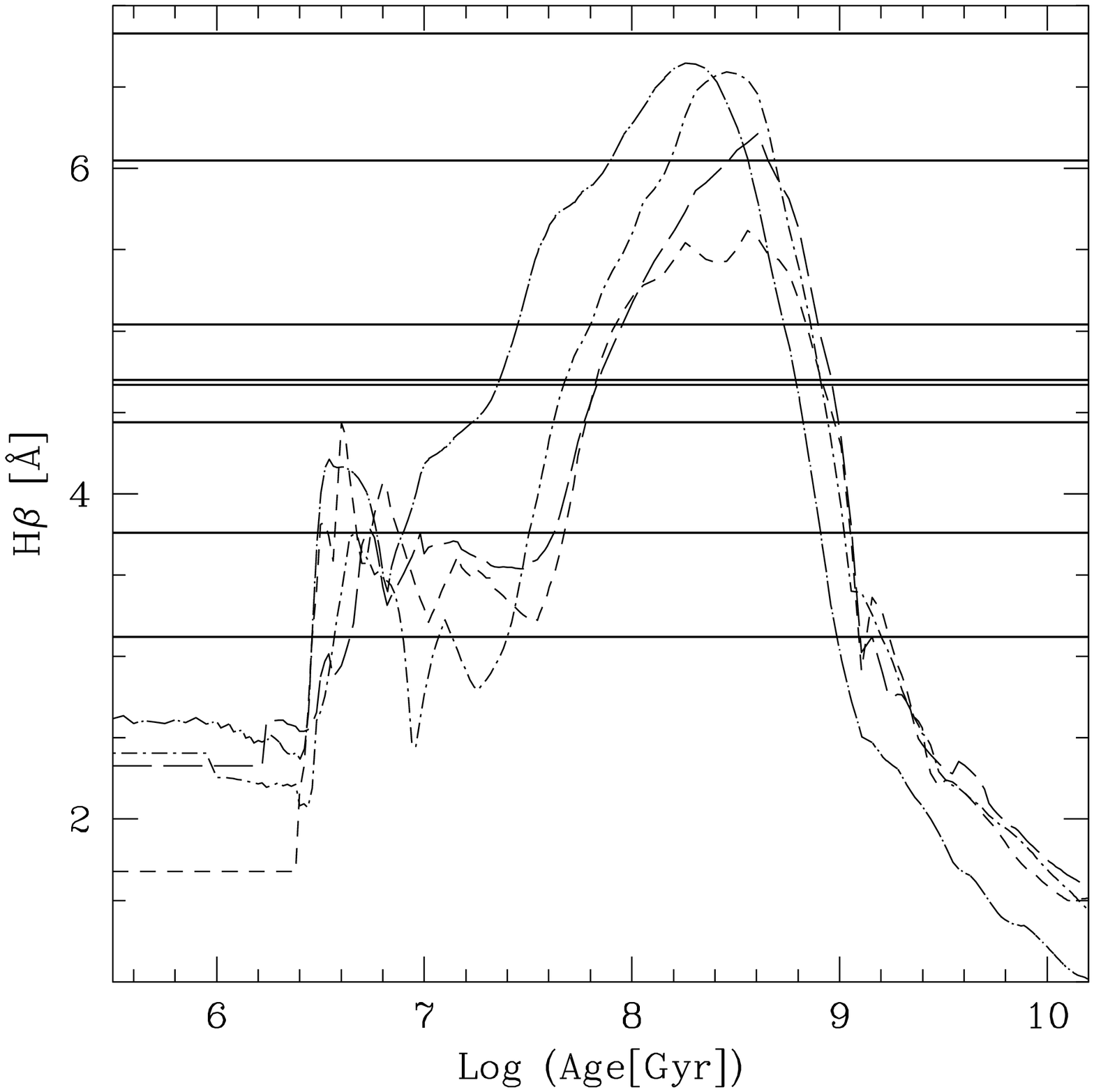}}
\figcaption{\small
Evolution of the Lick H$\beta$ index according to the
Bruzual \& Charlot (2003) evolutionary population
synthesis models, adopting the Chabrier (2003) IMF.
Curves in different line-types
represent metallicities $\Zsun$ ([Fe/H]): 0.004(--0.64), 0.008(--0.33), 
0.02(+0.09) and 0.05(+0.56) (short-dashed, 
long-dashed, dot-short dashed and dot-long dashed
lines respectively).
The thick horizontal lines are the H$\beta$ measurements
of the eight young clusters.
}
\label{BCplot}
\end{center}}

As has been noted before, the relatively narrow definition
of the Lick H$\beta$ index is not particularly 
well suited
for very young stellar populations (e.g., Larsen \& Brodie 2002).
However, we do not rely solely on the Balmer lines 
for age discrimination. Following the method
outlined in Proctor et al. (2004), we have 
determined the ages and metallicities
of the clusters by fitting all our measured Lick
indices to the predictions of the BC03 models.
Using the entire set of Lick indices maximizes 
the amount of age and metallicity information
present in the spectra, since each index 
is individually sensitive to age and metallicity 
(and chemical abundance ratio variations; Proctor \& Sansom 2002). 
This approach is particularly
useful in this case since in these young clusters the 
metal-lines are extremely diluted.

BC03 provide Lick indices 
for their model SEDs which cover the age-range
0.125 Myr -- 20 Gyr, and a metallicity range 
of 0.0001 -- 0.05 $\Zsun$. In this case, we
use Lick indices measured off the BC03 SEDs, 
not those derived using the Lick fitting functions
since our data is not corrected to the Lick system. 
As is the case for the Lick stellar library, the STELIB library
is based on solar neighborhood stars. 
As such, it 
reflects the solar neighborhood abundance pattern.
This creates an abundance ratio bias which
is implicit in stellar population models
which employ empirical stellar libraries
(Trager et al. 2000; Thomas \& Maraston 2002; 
Proctor \& Sansom 2002).
For very young systems correcting for this 
bias is highly uncertain, and therefore
we make no correction to the BC03 models. This 
should not introduce serious errors in our derived 
abundances near to solar metallicity (Proctor et al. 2004).

For each cluster, we perform a $\chi^2$ fit of all our
measured indices to each combination of age and metallicity
in the BC03 models. The age and metallicity of the 
model which is best found to fit these data are then adopted
as the best age and metallicity estimate of the particular 
cluster in question. For detailed discussions of the method applied
to galaxy spectra see Proctor \& Sansom (2002) and
Proctor et al. (2004). The results of our $\chi^2$
analysis are given in Table~\ref{ChiTab}, and we show
the 'age-metallicity' relation for these clusters
in Figure~12. For comparison, we have also performed
the same analysis for a subset 
of the Beasley et al. (2002) LMC clusters.

\vbox{
\begin{center}
\leavevmode
\hbox{%
\epsfxsize=9cm
\epsffile{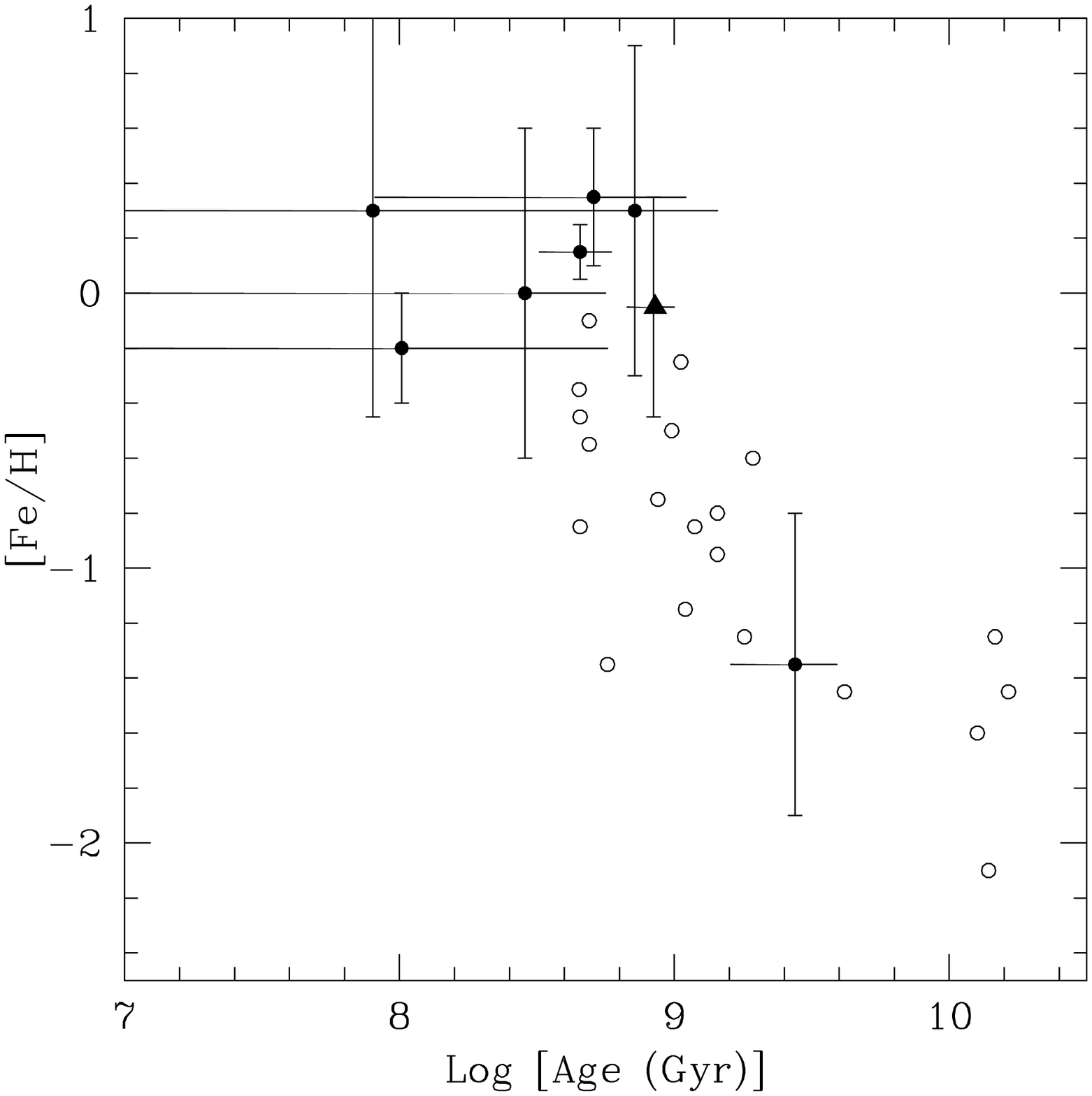}}
\figcaption{\small
Ages and metallicities for the young M31 clusters (filled
circles with error bars) and LMC clusters from
Beasley et al. (2002; open circles, error bars
omitted for clarity) based on the 
predictions of the Bruzual \& Charlot (2003) stellar
population models.
The position of Hubble V, the NGC~205 cluster, is marked
with a solid triangle. 
}
\label{AMR}
\end{center}}

The majority of the ages derived from the BC03 models 
are consistent with those obtained from the LMC 
cluster templates. 
However, the BC03 models 
do generally predict older ages than these
templates. This is consistent with the idea
that the LMC templates are somewhat more metal-poor
than the young M31 clusters.
For Hubble V, we obtain an 
age of 0.51$\pm$0.59 Gyr, which is consistent
with previous studies 
(Da Costa \& Mould 1988; Lee 1996).
Figure~12 shows that, with the exception
of 292-010, all of the young clusters are less
than 1 Gyr old, and are very close to solar metallicity.
Cluster 292-010 appears to be closer to 3 Gyr old, 
and is also somewhat more metal-poor. 
We conclude that our age determinations
are in reasonable accord with the LMC templates, 
and henceforth we adopt the age and metallicity estimates 
for the young clusters based on the BC03 models
which are given in Table~\ref{ChiTab}.
We are now in a position to 
estimate the masses of the clusters.

We have de-reddened the V-band apparent magnitudes
of the clusters using the mean
reddening, E(B--V)=0.22, derived by Barmby et al. (2000)
and adopting $R_V=3.2$ (Cardelli, Clayton \& Mathis 1989).
Assuming a distance modulus to M31 of 
$\mu$=24.47 (Holland 1998; Staneck \& Garnavich 1998), 
we obtain absolute magnitudes, and hence total 
V-band luminosities for the clusters.
$M/L_V$ ratios for the clusters come from BC03, 
based upon our BC03 age determinations. 
This has the benefit of
adding internal consistency to our calculations.
In Table~\ref{ChiTab} we list the adopted 
$M/L_V$ ratios and estimated masses
for the young M31 clusters.
Estimating the uncertainties in our mass determinations
is difficult, a major contribution to the error
budget is the unknown reddening in the disk
of M31. An increase in our adopted mean reddening 
by a factor of 2 (E(B--V)=0.44) would
roughly increase our mass estimate 
by the same factor. An adopted
factor of 2 uncertainty in the values given 
in Table~\ref{ChiTab} is not unreasonable.
The Chabrier (2003) IMF used in the 
BC03 models predicts that aging these
clusters to $\sim$13 Gyr will result
in their fading by up to four magnitudes.
Given their current absolute 
magnitude range of 
--8.57 $\leq M_V \leq$ --7.24, this
will result in clusters with 
--4.57 $\leq M_V \leq$ --3.24, which is 
some four magnitudes fainter than the peak of
the Galactic GC luminosity function.

A further consideration in this
mass calculation is the effect
of dynamical evolution, both internal
and external to the clusters (e.g., Fall \& Zhang 2001).
Several effects are expected to occur on different
time-scales which will act to lower the 
cluster masses. For ages $\leq 10^8$
years, stellar winds and supernovae will
remove some $\sim$ 30\% of the initial mass 
from a cluster, whereas for clusters older 
than this, two-body relaxation,  
bulge and/or disk shocking and dynamical friction
will all act upon the clusters to varying
degrees.  
The survival of Galactic {\it open} clusters (OCs) at a given 
mass depends crucially upon their linear size
(Wielen 1971).
Massive, but small clusters, will feel the effects
of dynamical relaxation most strongly, whereas
more extended clusters will be more susceptible 
to the effects of the galaxian tidal field. 
For the M31 clusters in question, 
with masses $\sim10^4\Msun$, we expect two-body
relaxation to be the primary 
source of mass evolution (Gnedin \& Ostriker 1997).
Obtaining structural 
parameters for these clusters will shed light
on this important issue. Suffice to say that our
mass estimates do not correct for any 
dynamical evolution.

The cluster masses and 'faded' 
absolute magnitudes we derive are relatively
modest in GC terms. Some 90\% of
the Galactic GC system have masses 
$\geq 1\times10^5\Msun$. The only cluster in the 
sample that lies in this GC mass regime is the
slightly older, more metal-poor
292-010, which lies out of the plane
of the disk (see later).
For reference, one of the least massive GCs known 
in the Galaxy is AM-4, which has an absolute 
magnitude, $M_V$=--1.24. 
Assuming $M/L_V$=2 for an old stellar population, we obtain a mass of 
$8.4\times10^3\Msun$ for this cluster.

An important question here is whether these
are actually young GCs, OCs, or lie somewhere else 
in the mass spectrum of star clusters.
After Larsen \& Richtler (1999), in Figure~13 we plot
the young M31 clusters in the $M_V$--age
plane compared to the Galactic OC
catalogue of Lyng\aa\ (1987), and the young, massive
clusters (YMCs) identified in a sample of
nearby spirals by Larsen \& Richtler (1999).

\vbox{
\begin{center}
\leavevmode
\hbox{%
\epsfxsize=9cm
\epsffile{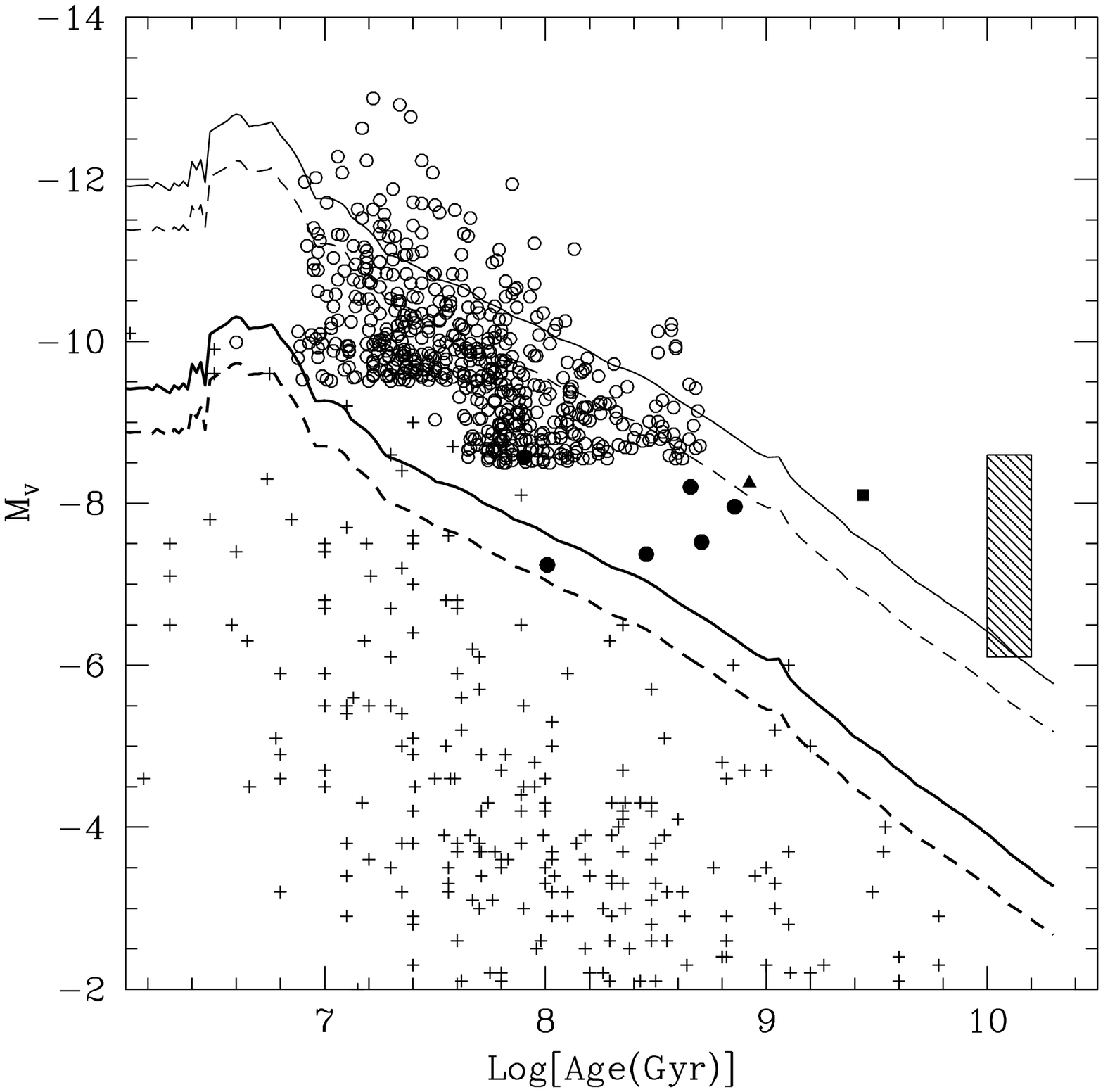}}
\figcaption{\small
Comparison of the young M31 clusters (solid symbols), open cluster data
from Lyng\aa\ (1987; plus signs) and young massive cluster data
from Larsen \& Richtler (1999; open circles) in the $M_V$--log(age)
plane. Hubble V in NGC~205 is marked as a triangle, 
292-010 which lies out of the plane of the disk 
is marked as a square. 
The hatched region denotes the area occupied by 
Milky Way GCs with mean $M_V$=--7.4, and a dispersion
of 1.2 mag.
Lines represent the predictions of the Bruzual \& Charlot (2003)
models for the fading with age of a cluster at masses
$M=10^4\Msun$ (thick lines) and $M=10^5\Msun$ (faint lines)
at solar metallicity.
Dashed and solid lines represent Salpeter and Chabrier IMFs
respectively.
}
\label{Lyngaa}
\end{center}}

The comparison between the young M31 clusters, the
OCs and YMCs is not strictly fair for two reasons. 
Firstly, ages are derived for these latter two cluster types
using integrated colors, whereas the M31 clusters
have spectroscopically derived ages. Secondly, and perhaps more
importantly, the selection functions of the three cluster types are
very different. 
However, Figure~13 is revealing. The Galactic
OCs from Lyng\aa\ (1987) are rather well bounded by
the BC03 models for $M=10^4\Msun$ at solar metallicity.
Very few are found to lie above this mass for the Chabrier
IMF. With the exception of 327-053, 
all the M31 clusters are well separated from the upper envelope
of the Galactic OCs. In fact, the M31 clusters roughly
cover the mass range of Larsen \& Richtler's (1999)
YMCs, although are somewhat older. This is probably
not a real effect, but reflects a bias in the
Larsen \& Richtler sample in that they preferentially
obtained data for the bright end of the YMC luminosity 
function, which are intrinsically younger YMCs.

Also in Figure~13 we show the position of a
GC at the turnover of the GCLF ($M_V=-7.4$), 
with an observed dispersion of 1.2 mag.
As the young M31 clusters fade and follow the 
locus of the models, they will occupy 
an area slighly below this region, between the 
GCs and OCs. Therefore, at present, we cannot
say for certain whether we are seeing massive 
OCs, or relatively low-mass GCs.
High-resolution imaging, and dynamical mass estimates
will distinguish between these two possibilities.
 
With the exceptions of Hubble V 
(an NGC~205 cluster) and 292-010,
all the clusters lie at or around
solar metallicities, are $<$ 1 Gyr old,
and are projected onto the disk
of M31 (Figure~\ref{Spatial}). 
Cluster 292-010 lies $\sim$10 kpc to 
the south-west of NGC~205, approximately
7 kpc away from the disk.
Are we, in fact,  looking at a disk 
system of young clusters in M31?
We first need to re-determine the velocities
of these young clusters prior
to looking at their kinematics, since their
broad Balmer lines lead to serious systematic
errors when their spectra are cross-correlated
against classical G and K templates (Barmby et al., 2000).
This possibility may afflict part of
the Perrett et al. (2002) sample.
We derive new velocities for the young
clusters by cross-correlation with a series
of A- and F- star templates from the STELIB
spectral library. These new velocities 
and their respective uncertainties are given in
Table~\ref{ChiTab}.
In Figure~14, we plot the
rotation curve of the six young clusters
compared to the rotation curve
derived from H{\small II} regions 
of M31 by Rubin \& Ford (1970), and
the H{\small I} curve of Kent (1989)

\vbox{
\begin{center}
\leavevmode
\hbox{%
\epsfxsize=9cm
\epsffile{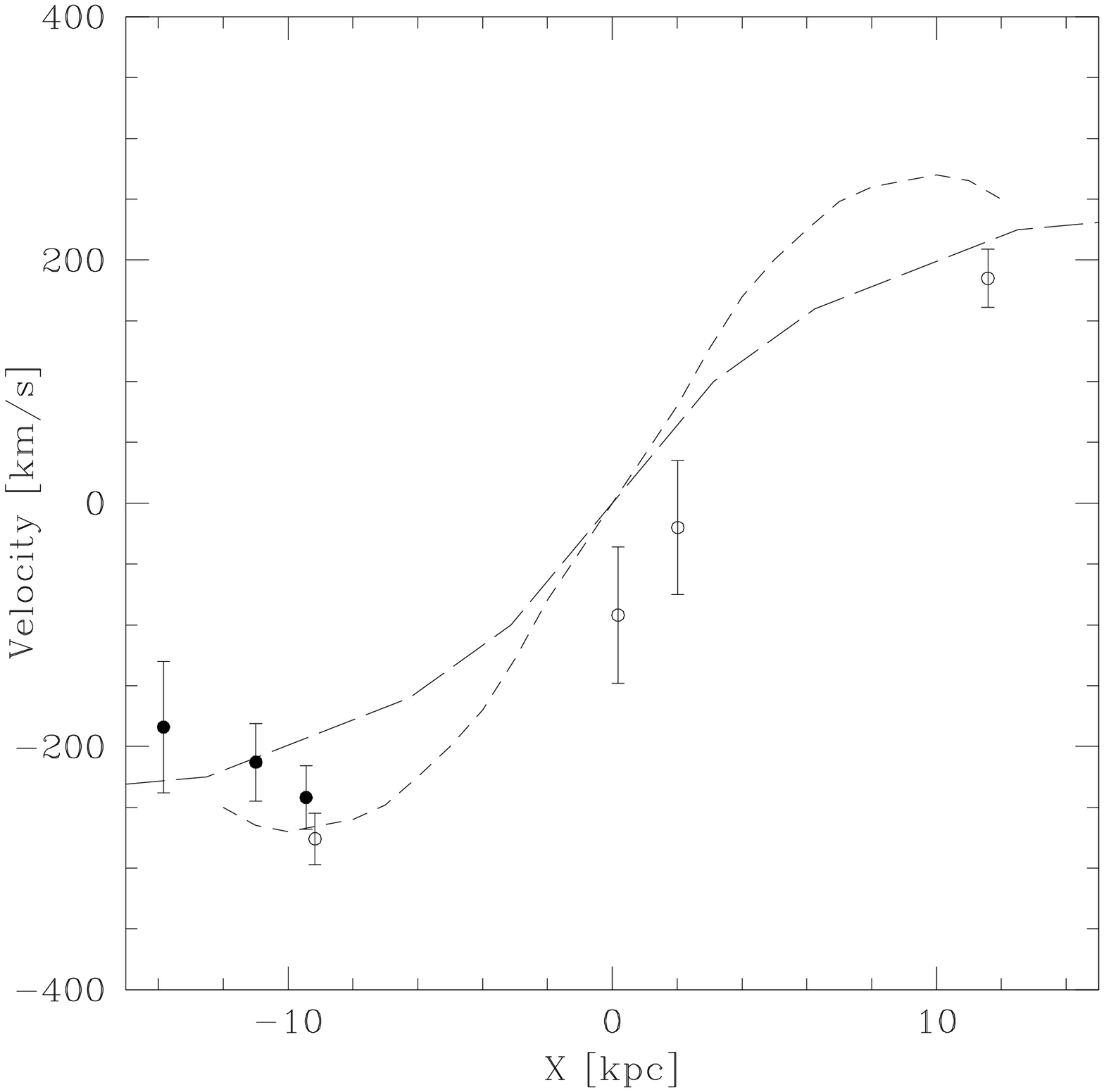}}
\figcaption{\small
Velocity versus the projected distance along the
major axis of M31. The six young clusters
are compared to the rotation of M31
H{\small II} regions from Rubin \& Ford (1970; short-dashed line) and the 
H{\small I} rotation curve of Kent (1989; long-dashed line)
assuming the same distance modulus.
The H{\small I} and H{\small II} data have been folded
on the $x$- and $y$-axes, and extrapolated
interior to 3 kpc. All velocities have been corrected to the 
M31 rest frame by +300 km/s.
Solid symbols represent the three clusters identified
as possessing thin disk kinematics by Morrison et al. (2004).
The young clusters show evidence of rotation
similar to that of M31's gas disk.
}
\label{Rotation}
\end{center}}

The clusters show signs of rotating 
in the same sense as, and with a similar
amplitude to, these H$\textsc{I}$ and H$\textsc{II}$ 
data. 
Unfortunately, the small sample and low-resolution
of our spectra are unsuitable for a detailed 
analysis of this issue.
Interestingly, on the basis of accurate velocities
for $\sim$200 M31 GCs (Perrett et al. 2002), 
Morrison et al. (2004) have suggested the 
existence of a {\it thin} disk system of globular 
clusters in M31.
This is in contrast to the Milky Way, which
seems to have no such system. 
Moreover, these authors showed that, based upon the Perrett et al. 
metallicities, 
the clusters cover a metallicity range --2.0$\leq$[Fe/H]$\leq$ 0. 
This is significantly more metal-poor than the Milky Way stellar thin disk.
Morrison et al. (2004) argued that, although they had no 
secure age estimates for these disk clusters, the fact that the 
disk clusters extend down to such low metallicities was indicative that 
the thin disk of M31 was likely to be
old ($\sim$ 10 Gyr).

Four of the six disk clusters in our sample are included in
the Morrison et al. (2004) study, namely 314-037, 321-046, 327-053
and 380-313. Of these four, three have residuals from a disk
model $\leq$ 0.75, corresponding to thin-disk kinematics
(see Morrison et al. 2004 for the precise 
definitions): 314-037(0.57),  321-046(--0.05)
and 327-053(0.25). These are consistent with their definition 
of belonging to a thin disk population, and indeed these clusters
lie near the rotation curves shown in Figure~14.
However, rather than being old, metal-poor disk clusters as suggested
by Morrison et al.(2003), we find that these clusters are 
young and at approximately solar metallicity.
We suggest that age-metallicity degeneracy, and the
lack of Balmer lines in the Perrett et al. (2002)
spectra, lead these authors to conclude incorrectly
these were metal-poor rather than young objects.
A full census of the ages of these disk clusters 
is required to investigate this point.

\section{Discussion}
\label{Discussion}

There appear to be at least two striking differences
between the M31 GC system and that of the Milky Way.

{\it (i). Nitrogen Enhancement}

Since the work of Burstein et al. (1984) it has
been suggested that CN in the integrated spectra
of M31 GCs is enhanced relative to Milky Way GCs at the
same metallicity.
We have shown that this remains true for GCs
in the range --1.5 $\leq$ [Fe/H] $\leq$ --0.3
(the metallicity range of the Milky Way GCs in our sample).
Moreover, both the M31 and Milky Way GCs are enhanced
in CN when compared to elliptical galaxies,  
which themselves show differences when divided into
field and cluster samples 
(S$\acute{a}$nchez-Bl$\acute{a}$zquez et al. 2003).
The very existence of CN variations
suggest than N is the main culprit, since
C is expected to be effectively absorbed into the CO
molecule. This is supported by observations
of the NH feature at $\lambda$3360~\AA\ (Ponder et al. 1998; 
Li \& Burstein 2004). The behaviour of C is less clear, 
although from the point of view of stellar evolution, 
C is expected to be destroyed in favor of N if
the sum of C+N+O is preserved. 
A better understanding of the behavior
of C and O in GCs may benefit from  a comprehensive 
study of the CO feature at 2.3$\mu$m (e.g., Davidge 1990).

Varying CN enhancements in the {\it integrated} spectra of
Galactic and M31 clusters is particularly interesting when one considers 
the CN variations known to exist between stars in individual Galactic
globulars. The best documented case is that of 47 Tucanae. 
Norris \& Freeman (1979) found that 
the giants in this cluster exhibited CN-strong and CN-weak bimodality
from the CN molecular band at $\lambda$4150~\AA. 
This bimodality has since been observed to several magnitudes
below the main sequence turn-off in Galactic GCs, 
along with the discovery that CN and CH are anti-correlated
as expected from the CNO cycle
(e.g., Cannon et al. 1998; Gratton et al. 2002; 
Briley, Cohen \& Stetson 2002, 2004; Harbeck, Smith \& Grebel 2003).
Although no such studies exist for the much more distant M31
GCs, Tripicco (1989) and Tripico \& Bell (1992), showed that, 
on the basis of near-UV 
integrated spectroscopy and synthetic spectra, CN enhancement in the M31 
GCs likely arises from CN-strong giants {\it and} CN-strong dwarfs.
Thus, the luminosity dependence of CN appears
to exist in M31 GCs as well as Milky Way clusters.

As discussed comprehensively by Cannon et al. (1998), and summarized
by Harbeck et al. (2003), three scenarios have been discussed
to explain these puzzling chemical inhomogeneities; 
internal evolution of GC stars, GC self-pollution from
evolved stars and primordial differences or GC self-enrichment.

Both the internal evolution and GC self-pollution
scenarios have difficulty explaining the observed 
bimodality in CN in general, and its existence well
below the main-sequence turn-off in GCs.
CNO variations are expected to occur on the giant branch from 
standard mixing theory, where 
the convective envelope reaches it closest penetration
to the CNO burning shell. However, this is not expected
in stars well below the main-sequence turn-off, 
which are not expect to undergo deep mixing (Renzini \& Voli 1981).
Self-pollution from asymptotic giant branch (AGB) stars
are expected to eject CNO products through stellar winds, 
which may be deposited upon the rest of the stars in 
the cluster (Ventura et al. 2001). However, 
such accreted material is expected to be diluted 
in red giant branch (RGB) stars which undergo deep
mixing. This is generally not observed, except possibly
in a few metal-poor GCs (Suntzeff 1981; Carbon et al. 1982).

The primordial scenario argues that the proto-globular
clouds had a range of chemical compositions, and 
a collision between two such clouds may result
in the observed bimodality, if the gas remained
unmixed prior to the onset of star formation.
This picture is not unreasonable 
in an early proto-galactic environment, when
the cross-section of cloud-cloud collisions 
is expected to be high (Bekki et al. 2004).
However, this scenario faces the constraint in that
the iron-peak elements show very little dispersion
in individual GCs (e.g. Kraft 1994).

All of the M31 GCs in our sample have similar
metallicities and luminosities to their
Galactic counterparts, and lie on the 
same fundamental plane (Barmby, Holland \& Huchra 2002).
These clusters are enhanced in CN with respect to the 
Milky Way GCs, which themselves show cluster to cluster 
variations (Cannon et al. 1998).
It seems unlikely that
internal-evolution or self-pollution would operate
preferentially in the M31 GCs, but not in 
Milky Way clusters, particularly since the M31 GC
dwarfs are likely CN-strong (Tripicco \& Bell 1992; 
Li \& Burstein 2004).
Unless there are significant age differences between
the Milky Way GCs and all the M31 clusters in our sample (which 
we have not observed, Paper II), it seems highly likely
that the source of the CN enhancement was external
to the M31 cluster system.

Although highly speculative, we are forced to
conclude that the nitrogen enhancement in
M31 GCs must be primordial in origin.
The M31 proto-globular clouds were 
presumably enriched in nitrogen at
some early epoch, whereas the 
proto-globular clouds of the Milky Way 
GCs were not. 
Much of the remaining
nitrogen-rich material must have become incorporated into the 
bulk of the galaxy itself, 
although the M31 bulge shows significantly 
lower CN than the M31 GCs.
With the advent of larger aperture telescopes, 
fine abundance analysis of individual halo 
stars in M31 would be extremely interesting
in this regard.

{\it (ii). Young Disk Clusters}

Six of the young clusters in our sample
appear to be disk clusters. They are projected
on to the M31 disk, appear to rotate with 
the H$\textsc{I}$ gas, and have approximately
solar metallicities. Interestingly, one non-disk cluster appears
to be somewhat older ($\sim$3 Gyr), and lies 
approximately 7 kpc away from the plane of the disk.
On the basis of our mass estimates, it
is presently not clear whether these objects
are true GCs. Provided that they survive the dual effects of the 
tidal field of M31, and two-body relaxation, 
they will possibly evolve to populate the fainter end of the GCLF, 
or populate the bright end of the OCLF.

Are these clusters unusual? 
The work of Larsen \& Richtler (1999, 2000) highlighted
the fact that young, massive clusters are actually a common
feature of local spiral galaxies. The
presence of young disk clusters in M31 is
therefore not surprising. 
Based on a sample of 31 nearby spirals, 
Larsen \& Richtler (2000) found a correlation
between the specific $U$-band YMC
luminosity, $T_L(U)$, and the star formation
rate (SFR) of the galaxies as traced by
the far infrared luminosity of the host galaxy.

It is interesting to ask the question, where does M31
lie on this relation? Using equation (7)
of Larsen \& Richtler (2000), based
upon the far-infrared luminosity of M31, we 
obtain a normalised SFR for this galaxy of:
$1.9\times10^{-4} \Msun yr^{-1} kpc^{-2}$.
Assuming that all the metal-poor ([Fe/H]$<$--1.0)
disk clusters in the Morrison et al. (2004)
sample are young clusters, and adopting
a mean $U$-band absolute magnitude of
$\langle M_U \rangle$=-6.7 (Barmby et al. 2000), 
we obtain $T_L(U)$=0.02.
Clearly the sample of M31 disk clusters is not 
complete, but these values suggest that M31 lies 
on the Larsen \& Richtler (2000) relation, and 
in the reasonably low SFR, low $T_L(U)$ regime 
(see figures 3 \& 5 of Larsen \& Richtler 2000).
This lends support to the idea that these 
objects are not open clusters in M31, 
and may represent the precursors of true
thin disk globular clusters.

A system of young disk globular clusters is not seen
in the Milky Way, and if M31 is truly forming disk
GCs then this has important implications
for galaxy formation theories. 
GCs are thought to trace the early episodes
of the collapse of their host galaxy (e.g., Searle \& Zinn 1978; 
Forbes, Brodie \& Grillmair 1997) as well as 
later accretion (Cot$\acute{e}$, Marzke \& West 1998)
and galaxy merging (Ashman \& Zepf 1992).
Our results suggest that GCs may also
trace the build up of galaxy disks, both at the present
epoch and at higher redshifts.
Indeed, the very existence
of such disk clusters is expected on the basis 
that disks are where the majority of present-day
star formation occurs. Recent efforts to 
simulate GC formation in the CDM paradigm
also identify that a large fraction of GCs
are expected to form in disk-like
structures at high redshift (Beasley et al. 2002; 
Kravtsov \& Gnedin 2003).

Morrison et al. (2004)
have recently argued for the existence 
of a thin disk in M31, based on GC kinematics, which 
extends down to [Fe/H]$\sim$--2.0.
Such an old disk is perhaps not unexpected from 
a theoretical standpoint, if it can avoid disruptive
merging in the dense, early universe (Baugh et al. 1996; 
Abadi et al. 2003). The observational constraints
in this regard are not so clear, since the stellar 
thin disk of the Milky Way is the only such disk 
heretofore known to exist. 
However, in general, spiral stellar disks are thought
to be metal-rich, if not relatively young (Bell \& de Jong 2002).

We have found that, rather than being old and
metal-poor, the M31 disk clusters are in fact young and metal-rich.
A larger census of these young disk clusters is required
to see if they populate the entire low-metallicity
sample of Morrison et al.(2004), however the implication is
clear. Based upon our cluster data, 
the thin-disk identified by Morrison et al. (2004)
may in fact be significantly younger than 10 Gyr
(possibly as young as the oldest $\sim$ 1 Gyr cluster), with a metallicity not 
too dissimilar to that of the Milky Way disk.

Brown et al. (2003) recently reported the presence
of a significant intermediate-aged (6--8 Gyr)
and metal-rich ([Fe/H] $>$--0.5) population
of stars in a minor-axis halo field of M31, 
based on deep HST/ACS imaging.
These authors suggested that M31 may have undergone
a nearly equal-mass merger, 6--8 Gyr in the past.
The presence of an old thin disk as identified by 
Morrison et al. (2004) effectively ruled out
such a scenario, since any disk would be disrupted 
in such an event.
With the identification that the thin disk of
M31 may be as young as $\sim$ 1 Gyr, the suggestion by 
Brown et al. (2003) becomes more tenable. 

Based upon the evidence of enhanced N, and a 
disk population of possible GCs in M31, 
it appears then that not all GC systems 
of spiral galaxies are born equal.
Both the initial chemical enrichment, 
and star formation (possibly merger history) of these
two galaxies have differed significantly.
Future work required includes a full census
of the young star clusters in M31, secure
mass determinations of these clusters, further near-UV
and IR spectroscopy to understand the role
of N and C in GCs respectively, and extensive
spectoscopic and NUV imaging surveys
of a greater number of local spiral galaxies.

\acknowledgments

We thank Javier Cenarro, Michael Pierce and Michael Rich
for useful comments and suggestions on the paper,  
and Chris Flynn and Brad Gibson for early discussions
regarding this work.
We also thank P. S$\acute{a}$nchez-Bl$\acute{a}$zquez
for supplying her thesis data ahead of full publication, 
S$\o$ren Larsen for prividing the YMC data.
We are also very grateful to Judy Cohen and John Blakeslee
for supplying the LRIS Galactic GC spectra in digital form.
Funding support comes from NSF grant AST 0206139.
The data presented herein were obtained at the
W.M. Keck Observatory, which is operated as a scientific partnership among
the California Institute of Technology, the University of California and
the National Aeronautics and Space Administration.  The Observatory was
made possible by the generous financial support of the W.M. Keck
Foundation. This research has made use of the NASA/IPAC Extragalactic
Database  (NED), which is operated by the Jet Propulsion Laboratory,
Caltech, under contract with the National Aeronautics and Space
Administration.

%% Generally speaking, only the figure captions, and not the figures
%% themselves, are included in electronic manuscript submissions.
%% Use \figcaption to format your figure captions. They should begin on a
%% new page.

%% No more than seven \figcaption commands are allowed per page,
%% so if you have more than seven captions, insert a \clearpage

%% There must be a \figcaption command for each legend. Key the text of the
%% legend and the optional \label in curly braces. If you wish, you may
%% include the name of the corresponding figure file in square brackets.
%% The label is for identification purposes only. It will not insert the
%% figures themselves into the document.
%% If you want to include your art in the paper, use \plotone.
%% Refer to the on-line documentation for details.

%\figcaption[sgi9259.eps]{This is the first figure and it uses sgi9259.eps as
%its EPS figure file. \label{fig1}}

%% Tables should be submitted one per page, so put a \clearpage before

%% Two options are available to the author for producing tables:  the
%% deluxetable environment provided by the AASTeX package or the LaTeX
%% table environment.  Use of deluxetable is preferred.
%%

%% Three table samples follow, two marked up in the deluxetable environment,
%% one marked up as a LaTeX table.

%% In this first example, note that the \footnotesize command has been
%% used to shrink the table so it will fit on one page. Note also that
%% the \label command needs to be placed inside the \tablecaption.

\clearpage

\begin{deluxetable}{lrrrrrrrrl}
\renewcommand{\arraystretch}{.6} 
\rotate
\footnotesize
\tablecaption{Basic data for M31 globular clusters. \label{BasicData}}
\tablewidth{0pt}
\tablehead{
\colhead{Name$^a$} & \colhead{RA$^b$}   & \colhead{Dec$^b$}   &
\colhead{X$^b$} &
\colhead{Y$^b$}  & \colhead{Z$^b$} & \colhead{Velocity$^b$} & 
\colhead{S/N$^c$} & \colhead{Notes} \\
\colhead{} & \colhead{(J2000)}   & \colhead{(J2000)}   &
\colhead{(arcmin)} &
\colhead{(arcmin)}  & \colhead{(arcmin)} & \colhead{(kms$^{-1}$)} & 
\colhead{(pixel$^{-1}$)} & &
}
\startdata
126-184 &   0 42 43.78 &  41 12 42.98 &  -2.8 &  -2.0 &   3.5 & --182$\pm$14 & 73 &\\
134-190 &   0 42 51.72 &  41 14 02.88 &  -0.8 &  -2.4 &   2.5 & --401$\pm$32 & 70 &\\
158-213 &   0 43 14.47 &  41 07 20.58 &  -3.4 &  -9.9 &  10.5 & --187$\pm$6$^d$ & 120 & velocity standard\\
163-217 &   0 43 17.71 &  41 27 45.53 &  13.0 &   2.2 &  13.2 & --161$\pm$4$^d$ & 170 & velocity standard\\
222-277 &   0 44 25.51 &  41 14 12.62 &  10.2 & -16.2 &  19.1 & --282$\pm$20& 30 & F-type spectrum, peculiar\\
225-280 &   0 44 29.78 &  41 21 36.57 &  16.5 & -12.2 &  20.5 & --167$\pm$8$^d$ & 169 & velocity standard\\
234-290 &   0 44 46.50 &  41 29 18.34 &  24.5 &  -9.9 &  26.4 & --234$\pm$16& 71 &\\
292-010 &   0 36 16.69 &  40 58 26.64 & -58.5 &  47.1 &  75.1 & --331$\pm$23& 32 &late F spectrum\\
301-022 &   0 38 21.61 &  40 03 37.22 & -87.8 &  -5.0 &  88.0 & --30$\pm$20 & 31 &\\
304-028 &   0 38 56.82 &  41 10 28.81 & -30.7 &  30.3 &  43.1 & --353$\pm$15& 34&\\
310-032 &   0 39 25.82 &  41 23 33.46 & -17.0 &  34.0 &  38.0 & --189$\pm$18& 35&\\
313-036 &   0 39 44.52 &  40 52 56.24 & -39.1 &  12.5 &  41.1 & --446$\pm$6 & 52&\\
314-037 &   0 39 44.60 &  40 14 08.23 & -69.9 & -11.2 &  70.7 & --318$\pm$66& 17 & F0 spectrum\\
321-046 &   0 40 15.30 &  40 27 46.86 & -55.5 &  -7.5 &  56.0 & --527$\pm$35& 30 & A-type spectrum\\
322-049 &   0 40 17.28 &  40 39 05.83 & -46.3 &  -0.8 &  46.3 & --581$\pm$29& 40 & A-type spectrum\\
324-051 &   0 40 20.75 &  41 40 50.79 &   3.0 &  36.4 &  36.5 & --299$\pm$35& 69 & F-type spectrum, Hubble V\\
327-053 &   0 40 24.09 &  40 36 22.75 & -47.7 &  -3.5 &  47.8 & --658$\pm$29& 71 & A-F spectrum\\
328-054 &   0 40 24.86 &  41 40 23.74 &   3.1 &  35.5 &  35.6 & --243$\pm$23& 30&\\
337-068 &   0 40 48.45 &  42 12 10.45 &  30.9 &  51.5 &  60.0 & 50$\pm$12& 178&CN weak, CH strong?\\
347-154 &   0 42 22.95 &  41 54 28.25 &  27.7 &  26.7 &  38.5 & --312$\pm$36& 61&\\
350-162 &   0 42 28.43 &  40 24 52.18 & -42.2 & -29.2 &  51.4 & --457$\pm$18& 39&\\
365-284 &   0 44 36.77 &  42 17 20.47 &  61.0 &  21.4 &  64.7 & --78$\pm$13& 60&\\
380-313 &   0 46 06.34 &  42 00 53.23 &  58.5 &  -1.9 &  58.5 & --121$\pm$31& 43 &F0 spectrum\\
383-318 &   0 46 11.89 &  41 19 42.15 &  27.0 & -28.4 &  39.1 & --250$\pm$10& 85&\\
393-330 &   0 47 01.22 &  41 24 06.45 &  36.2 & -32.8 &  48.9 & --331$\pm$10& 43&\\
398-341 &   0 47 57.83 &  41 48 44.63 &  62.0 & -25.6 &  67.1 & --227$\pm$5& 44&\\
401-344 &   0 48 08.52 &  41 40 42.47 &  57.0 & -32.2 &  65.5 & --273$\pm$52& 54&\\
NB16    &   0 42 33.18 &  41 20 17.12 &   2.0 &   4.2 &   4.6 & --115$\pm$15&142&\\
NB67    &   0 42 34.18 &  41 19 47.10 &   1.7 &   3.7 &   4.1 & --113$\pm$17& 91&\\
NB89    &   0 42 44.69 &  41 14 41.97 &  -1.1 &  -1.0 &   1.5 & --332$\pm$6 & 110&\\
\enddata
%% Text for table notes should follow after the \enddata but before
%% the \end{deluxetable}. Make sure there is at least one \tablenotemark
%% in the table for each \tablenotetext.

\tablenotetext{a}{Nomenclature from Brodie, Huchra \& Kent (1991).}
\tablenotetext{b}{From Barmby et al. (2000).}
\tablenotetext{c}{Derived signal-to-noise measured at 5000\AA.}
\tablenotetext{d}{Echelle velocities from Peterson (2000)}

\tablecomments{X,Y,Z are in arcmins and defined in Huchra, Brodie \& Kent (1991)
}

\end{deluxetable}

\clearpage

%\begin{turnpage}
\begin{deluxetable}{lrrrrrrrrrrrrrr}
\renewcommand{\arraystretch}{.6} 
\rotate
\footnotesize
\tablecaption{Lick/IDS indices and uncertainties measured for the Galactic GCs
in the Cohen et al.(1998) sample \label{CBR98Indices}}
\tablewidth{0pt}
\tablehead{
\colhead{Name} & \colhead{H$\delta_{\rm A}$}   & \colhead{H$\delta_{\rm F}$}   &
\colhead{CN$_1$} & \colhead{CN$_2$}  & \colhead{Ca4227} &
\colhead{G4300} & \colhead{H$\gamma_{\rm A}$}  & \colhead{H$\gamma_{\rm F}$} &
\colhead{Fe4384} & \colhead{Ca4455}  & \colhead{Fe4531} &
\colhead{C$_2$4668}  & \colhead{H$\beta$}  & \colhead{Fe5015} 
%\colhead{Mg$_1$} \colhead{Mg$_2$}  & \colhead{Mg $b$}  & \colhead{Fe5270} &
%\colhead{Fe5335} & \colhead{Fe5406} &
%\colhead{Fe5709} \colhead{Fe5782}  & \colhead{Na D}  & \colhead{TiO$_1$} &
%\colhead{TiO$_2$} 
}
\startdata
NGC~6205 & ... &... & ... & ... &  0.32 &  2.14 &  1.54 &  1.90 &  0.86 &  0.11 &  1.10 &  0.19 &  2.47 &  1.43\\ 
... & ... &... & ... & ... &  0.03 &  0.57 &  0.54 &  0.17 &  0.34 &  0.05 &  0.07 &  0.46 &  0.13 &  0.23\\ 
NGC~6121 & ... &... & ... & ... &  0.40 &  2.72 &  1.58 &  2.23 &  1.76 &  0.22 &  1.48 &  0.62 &  2.91 &  2.26\\ 
... & ... &... & ... & ... &  0.05 &  0.58 &  0.55 &  0.18 &  0.35 &  0.07 &  0.10 &  0.47 &  0.13 &  0.24\\ 
NGC~6838 & ... &--0.34 & 0.005 & 0.023 &  0.81 &  5.30 & --4.57 & --1.11 &  3.13 &  0.57 &  2.55 &  2.10 &  1.43 &  3.79\\ 
... & ... & 0.12 & 0.003 & 0.003 &  0.03 &  0.57 &  0.54 &  0.17 &  0.34 &  0.04 &  0.06 &  0.46 &  0.13 &  0.22\\ 
NGC~6341 & ... &... & ... & ... &  0.17 &  1.04 &  2.78 &  2.44 &  0.61 &  0.09 &  0.57 &  0.08 &  2.55 &  0.66\\ 
... & ... &... & ... & ... &  0.03 &  0.57 &  0.54 &  0.17 &  0.34 &  0.04 &  0.07 &  0.46 &  0.13 &  0.23\\ 
NGC~6171 & ... &... & ... & ... & ... & --0.47 &  4.96 &  2.52 &  1.48 &  0.14 & --0.51 &  0.32 &  1.60 &  2.53\\ 
... & ... &... & ... & ... & ... &  0.61 &  0.57 &  0.21 &  0.46 &  0.16 &  0.23 &  0.54 &  0.16 &  0.30\\ 
NGC~6356 & ... & 0.60 & 0.044 & 0.067 &  0.75 &  4.97 & --3.81 & --0.60 &  3.25 &  0.69 &  2.62 &  2.06 &  1.63 &  3.57\\ 
... & ... & 0.13 & 0.003 & 0.003 &  0.04 &  0.57 &  0.55 &  0.18 &  0.34 &  0.05 &  0.08 &  0.46 &  0.13 &  0.23\\ 
NGC~6440 & ... &... & ... & ... & ... & ... & ... & ... & ... & ... & ... &  2.42 &  1.44 &  3.90\\ 
... & ... &... & ... & ... & ... & ... & ... & ... & ... & ... & ... &  0.46 &  0.13 &  0.23\\ 
NGC~6528 & ... &... & ... & ... & ... & ... & ... & ... & ... & ... & ... &  3.88 &  1.83 &  5.00\\ 
... & ... &... & ... & ... & ... & ... & ... & ... & ... & ... & ... &  0.47 &  0.13 &  0.23\\ 
NGC~6539 & ... &... & ... & ... & ... & ... & ... & ... & ... & ... & ... &  2.68 &  1.40 &  3.94\\ 
... & ... &... & ... & ... & ... & ... & ... & ... & ... & ... & ... &  0.51 &  0.15 &  0.27\\ 
NGC~6553 & ... &... & ... & ... & ... & ... & ... & ... & ... & ... & ... &  3.92 &  1.68 &  5.08\\ 
... & ... &... & ... & ... & ... & ... & ... & ... & ... & ... & ... &  0.46 &  0.13 &  0.22\\ 
NGC~6624 & ... &... & ... & ... & ... & ... & ... & ... & ... & ... & ... &  1.83 &  1.71 &  3.61\\ 
... & ... &... & ... & ... & ... & ... & ... & ... & ... & ... & ... &  0.45 &  0.13 &  0.22\\ 
NGC~6760 & ... &... & ... & ... & ... & ... & ... & ... & ... & ... & ... &  2.15 &  1.38 &  4.11\\ 
... & ... &... & ... & ... & ... & ... & ... & ... & ... & ... & ... &  0.47 &  0.13 &  0.23\\ 

\enddata

\end{deluxetable}
%\end{turnpage}

\clearpage

\begin{deluxetable}{lrrrrrrrrrr} 
\renewcommand{\arraystretch}{.6} 
\rotate
\footnotesize
\tablecaption{Lick/IDS indices and uncertainties measured for the Galactic GCs
in the Cohen et al.(1998) sample.}
\tablenum{2A}
\tablewidth{0pt}
\tablehead{
\colhead{Name} & \colhead{Mg$_1$} & \colhead{Mg$_2$}  & \colhead{Mg $b$}  & \colhead{Fe5270} &
\colhead{Fe5335} & \colhead{Fe5406} & \colhead{Fe5709} & \colhead{Fe5782}  & \colhead{Na D}  & 
\colhead{TiO$_1$}\\
\colhead{} & \colhead{[mag]} &  \colhead{[mag]} & \colhead{[\AA]} & \colhead{[\AA]} &  
\colhead{[\AA]} & \colhead{[\AA]} &  \colhead{[\AA]} & \colhead{[\AA]} & \colhead{[\AA]} &  
\colhead{[mag]}\\
}
\startdata
NGC~6205 & 0.008 &0.048 &  0.83 &  0.87 &  0.74 &  0.38 &  0.23 &  0.11 &  0.91 & 0.003\\ 
... & 0.014 &0.010 &  0.08 &  0.14 &  0.13 &  0.17 &  0.11 &  0.13 &  0.10 & 0.012\\ 
NGC~6121 & 0.031 &0.094 &  1.57 &  1.26 &  0.96 &  0.51 &  0.47 &  0.36 &  1.68 & 0.009\\ 
... & 0.014 &0.010 &  0.08 &  0.14 &  0.13 &  0.17 &  0.11 &  0.13 &  0.10 & 0.012\\ 
NGC~6838 & 0.074 &0.164 &  2.71 &  2.17 &  1.95 &  1.32 &  0.78 &  0.57 &  2.28 & 0.023\\ 
... & 0.014 &0.010 &  0.08 &  0.13 &  0.13 &  0.17 &  0.11 &  0.13 &  0.10 & 0.012\\ 
NGC~6341 & 0.004 &0.029 &  0.52 &  0.48 &  0.43 &  0.18 &  0.09 &  0.04 &  0.68 & 0.003\\ 
.... & 0.014 &0.010 &  0.08 &  0.13 &  0.13 &  0.17 &  0.11 &  0.13 &  0.10 & 0.012\\ 
NGC~6171 & 0.041 &0.118 &  1.81 &  1.50 &  1.21 &  0.68 &  0.49 &  0.47 &  1.86 & 0.017\\ 
. & 0.014 &0.010 &  0.12 &  0.17 &  0.17 &  0.19 &  0.12 &  0.14 &  0.12 & 0.012\\ 
NGC~6356 & 0.072 &0.172 &  3.02 &  2.01 &  1.84 &  1.21 &  0.79 &  0.65 &  3.05 & 0.030\\ 
... & 0.014 &0.010 &  0.08 &  0.14 &  0.13 &  0.17 &  0.11 &  0.13 &  0.10 & 0.012\\ 
NGC~6440 & 0.102 &0.225 &  3.25 &  2.32 &  1.73 &  1.19 &  0.90 &  1.02 &  4.21 & 0.030\\ 
... & 0.014 &0.010 &  0.08 &  0.13 &  0.13 &  0.17 &  0.11 &  0.12 &  0.10 & 0.012\\ 
NGC~6528 & 0.097 &0.246 &  3.81 &  2.79 &  2.21 &  1.62 &  1.05 &  1.00 &  5.12 & 0.044\\ 
... & 0.014 &0.010 &  0.08 &  0.14 &  0.13 &  0.17 &  0.11 &  0.13 &  0.10 & 0.012\\ 
NGC~6539 & 0.084 &0.193 &  2.81 &  2.14 &  1.54 &  1.19 &  0.92 &  0.96 &  4.05 & 0.037\\ 
... & 0.014 &0.010 &  0.10 &  0.15 &  0.15 &  0.18 &  0.11 &  0.13 &  0.11 & 0.012\\ 
NGC~6553 & 0.106 &0.262 &  3.99 &  2.96 &  2.36 &  1.67 &  0.97 &  1.09 &  3.70 & 0.048\\ 
... & 0.014 &0.010 &  0.08 &  0.13 &  0.13 &  0.17 &  0.11 &  0.12 &  0.10 & 0.012\\ 
NGC~6624 & 0.048 &0.161 &  2.83 &  1.93 &  1.59 &  1.07 &  0.65 &  0.57 &  2.47 & 0.031\\ 
... & 0.014 &0.010 &  0.07 &  0.13 &  0.12 &  0.17 &  0.10 &  0.12 &  0.10 & 0.012\\ 
NGC~6760 & 0.095 &0.225 &  3.30 &  2.30 &  1.78 &  1.29 &  0.83 &  0.87 &  3.51 & 0.045\\ 
... & 0.014 &0.010 &  0.08 &  0.14 &  0.13 &  0.17 &  0.11 &  0.13 &  0.10 & 0.012\\ 

\enddata
\end{deluxetable}

\clearpage

%%%%GalacticOffsets

\begin{deluxetable}{lcc}
\footnotesize
\tablecaption{Mean offsets and standard deviations ($\sigma_{\rm rms}$)
between P02 and CBR98 Indices at the Lick/IDS resolution, but
uncalibrated to the Lick system. \label{GalacticOffsets}}
\tablewidth{0pt}
\tablehead{
\colhead{Index} & \colhead{P02--CBR98}   & \colhead{$\sigma_{\rm rms}$} \\
}
\startdata
H$\delta_{\rm A}$	& ...	& ...	\\
H$\delta_{\rm F}$	& 0.1144	& 0.1144$^a$	\\
CN$_1$ 			& 0.0027 & 0.0027$^a$ \\
CN$_2$			& 0.0018 & 0.0018$^a$ \\
Ca4227			& 0.0044	& 0.0044$^a$	\\
G4300			& --0.5746	& 0.5746$^a$	\\
H$\gamma_{\rm A}$	&--0.5411	& 0.5411$^a$	\\
H$\gamma_{\rm F}$	&--0.1702	& 0.1702$^a$	\\
Fe4383			&--0.3277	& 0.3277$^a$	\\
Ca4455			&0.0042	&0.0042$^a$	\\
Fe4531			&0.0299	&0.0299$^a$	\\
C$_2$4668		&0.5646	&0.4533	\\
H$\beta$		&--0.0027&0.1257	\\
Fe5015			&--0.8856&0.2142	\\
Mg$_1$			&--0.0087	&0.0136	\\
Mg$_2$			&--0.0043	&0.0096	\\
Mg $b$			&--0.1616	&0.0723	\\
Fe5270			&0.0650	&0.1309	\\
Fe5335			&0.0025	&0.1199	\\
Fe5406			&--0.1590&0.1689	\\
Fe5709			&0.3026	&0.1036	\\
Fe5782			&0.0063	&0.1235	\\
NaD			&--0.1975&0.0979	\\
TiO$_1$			&0.0076	&0.0115	\\
TiO$_2$			&0.0228 &0.0123 \\
\enddata
%% Text for table notes should follow after the \enddata but before
%% the \end{deluxetable}. Make sure there is at least one \tablenotemark
%% in the table for each \tablenotetext.

\tablenotetext{a}{$\sigma_{\rm rms}$ taken to be the absolute difference.}

\end{deluxetable}

\clearpage

%\begin{turnpage}
\begin{deluxetable}{lrrrrrrrrrrrrrr}
\renewcommand{\arraystretch}{.6} 
\rotate
\footnotesize
\tablecaption{Lick/IDS indices and uncertainties for M31 globular clusters.
\label{M31Indices}}
\tablewidth{0pt}
\tablehead{
\colhead{Name} & \colhead{H$\delta_{\rm A}$}   & \colhead{H$\delta_{\rm F}$}   &
\colhead{CN$_1$} & \colhead{CN$_2$}  & \colhead{Ca4227} &
\colhead{G4300} & \colhead{H$\gamma_{\rm A}$}  & \colhead{H$\gamma_{\rm F}$} &
\colhead{Fe4384} & \colhead{Ca4455}  & \colhead{Fe4531} &
\colhead{C$_2$4668}  & \colhead{H$\beta$}  & \colhead{Fe5015} \\
\colhead{} & \colhead{[\AA]} &  \colhead{[\AA]} & \colhead{[mag]} &  \colhead{[mag]} &
\colhead{[\AA]} & \colhead{[\AA]} &  \colhead{[\AA]} & \colhead{[\AA]} &  \colhead{[\AA]} &
\colhead{[\AA]} & \colhead{[\AA]} &  \colhead{[\AA]} & \colhead{[\AA]} &  \colhead{[\AA]} \\
%\colhead{Mg$_1$} \colhead{Mg$_2$}  & \colhead{Mg $b$}  & \colhead{Fe5270} &
%\colhead{Fe5335} & \colhead{Fe5406} &
%\colhead{Fe5709} \colhead{Fe5782}  & \colhead{Na D}  & \colhead{TiO$_1$} &
%\colhead{TiO$_2$} 
}
\startdata
126-184 &  3.82 & 3.03 & --0.060 & --0.029 &  0.42 &  2.44 &  1.80 &  2.77 &  1.06 &  0.26 &  1.43 &  0.75 &  3.63 &  2.29\\ 
... &  0.53 & 0.47 & 0.017 & 0.018 &  0.17 &  0.30 &  0.29 &  0.17 &  0.50 &  0.21 &  0.31 &  0.44 &  0.14 &  0.37\\ 
134-190 &  0.58 & 0.91 & 0.010 & 0.040 &  0.64 &  4.53 & --2.57 &  0.34 &  2.86 &  0.49 &  2.38 &  1.73 &  1.76 &  2.85\\ 
... &  0.52 & 0.47 & 0.017 & 0.018 &  0.17 &  0.28 &  0.31 &  0.19 &  0.50 &  0.21 &  0.31 &  0.45 &  0.16 &  0.38\\ 
158-213 &  0.27 & 1.10 & 0.040 & 0.068 &  0.62 &  4.06 & --2.53 &  0.18 &  2.91 &  0.63 &  2.40 &  1.97 &  1.72 &  3.23\\ 
... &  0.44 & 0.38 & 0.015 & 0.016 &  0.09 &  0.16 &  0.19 &  0.12 &  0.37 &  0.12 &  0.19 &  0.28 &  0.10 &  0.27\\ 
163-217 & --0.40 & 0.68 & 0.160 & 0.207 &  1.03 &  5.06 & --5.26 & --1.02 &  4.52 &  1.25 &  3.37 &  4.06 &  1.72 &  4.79\\ 
... &  0.43 & 0.37 & 0.014 & 0.015 &  0.07 &  0.12 &  0.16 &  0.09 &  0.33 &  0.09 &  0.15 &  0.21 &  0.07 &  0.22\\ 
222-277 &  6.98 & 4.60 & --0.120 & --0.076 &  0.72 &  1.87 &  4.06 &  3.57 &  1.86 &  0.78 &  2.74 &  1.39 &  4.44 &  4.10\\ 
... &  0.67 & 0.62 & 0.022 & 0.025 &  0.31 &  0.57 &  0.51 &  0.31 &  0.85 &  0.40 &  0.59 &  0.89 &  0.31 &  0.72\\ 
225-280 & --0.88 & 0.64 & 0.090 & 0.129 &  0.78 &  4.57 & --3.71 & --0.22 &  3.82 &  0.99 &  2.84 &  2.95 &  1.81 &  4.25\\ 
... &  0.43 & 0.37 & 0.014 & 0.015 &  0.07 &  0.12 &  0.16 &  0.09 &  0.34 &  0.09 &  0.16 &  0.22 &  0.07 &  0.23\\ 
234-290 & --0.10 & 0.85 & 0.000 & 0.030 &  0.75 &  4.37 & --3.21 & --0.18 &  2.49 &  0.52 &  2.56 &  1.77 &  1.70 &  3.14\\ 
... &  0.51 & 0.45 & 0.016 & 0.018 &  0.15 &  0.26 &  0.30 &  0.18 &  0.48 &  0.20 &  0.30 &  0.44 &  0.16 &  0.38\\ 
292-010 &  4.47 & 3.39 & --0.070 & --0.028 &  0.29 &  2.45 &  1.79 &  2.55 &  0.82 &  0.14 &  0.66 &  1.23 &  3.12 &  1.67\\ 
... &  0.64 & 0.58 & 0.021 & 0.023 &  0.29 &  0.51 &  0.50 &  0.30 &  0.82 &  0.39 &  0.60 &  0.88 &  0.32 &  0.76\\ 
301-022 &  3.60 & 2.17 & --0.080 & --0.056 &  0.89 &  3.02 &  0.85 &  1.91 &  1.21 &  0.43 &  1.73 &  0.72 &  2.92 &  2.26\\ 
... &  0.78 & 0.76 & 0.024 & 0.027 &  0.35 &  0.63 &  0.61 &  0.38 &  0.94 &  0.44 &  0.65 &  0.95 &  0.33 &  0.74\\ 
%302-02 &  2.24 & 2.25 & --0.070 & --0.041 &  0.50 &  2.12 &  1.70 &  1.79 &  0.34 &  0.09 &  1.70 &  1.25 &  2.05 &  2.16\\ 
%302-02 &  0.81 & 0.75 & 0.024 & 0.027 &  0.38 &  0.67 &  0.64 &  0.40 &  1.04 &  0.50 &  0.74 &  1.10 &  0.40 &  0.91\\ 
304-028 &  1.94 & 2.05 & --0.050 & --0.019 &  0.32 &  2.29 &  1.13 &  1.87 &  0.57 &  0.23 &  0.68 & --0.18 &  2.50 &  0.59\\ 
... &  0.69 & 0.64 & 0.021 & 0.024 &  0.30 &  0.53 &  0.51 &  0.32 &  0.82 &  0.39 &  0.59 &  0.86 &  0.30 &  0.72\\ 
%305-D0 &  1.03 & 1.85 & --0.000 & 0.041 &  0.66 &  2.90 & --0.76 &  2.45 &  4.26 &  1.24 &  1.85 &  2.34 &  4.73 &  2.17\\ 
%305-D0 &  1.07 & 1.00 & 0.031 & 0.036 &  0.51 &  0.92 &  0.94 &  0.54 &  1.32 &  0.67 &  1.04 &  1.54 &  0.54 &  1.31\\ 
%307-03 &  4.29 & 3.25 & --0.060 & --0.032 &  1.18 &  3.62 & --0.19 &  1.79 &  2.79 &  0.60 &  3.74 &  3.10 &  3.76 &  3.87\\ 
%307-03 &  0.85 & 0.80 & 0.027 & 0.031 &  0.40 &  0.72 &  0.74 &  0.45 &  1.10 &  0.54 &  0.77 &  1.18 &  0.43 &  1.00\\ 
310-032 &  4.07 & 2.98 & --0.110 & --0.072 &  0.44 &  1.77 &  1.34 &  1.85 &  0.62 &  0.01 &  1.26 & --0.22 &  2.54 &  1.45\\ 
... &  0.68 & 0.64 & 0.021 & 0.024 &  0.31 &  0.55 &  0.52 &  0.32 &  0.84 &  0.40 &  0.59 &  0.86 &  0.30 &  0.70\\ 
313-036 &  0.51 & 0.96 & 0.010 & 0.046 &  0.75 &  4.20 & --2.75 & --0.01 &  2.53 &  0.60 &  2.14 &  2.09 &  1.49 &  3.00\\ 
... &  0.56 & 0.51 & 0.017 & 0.019 &  0.20 &  0.34 &  0.37 &  0.23 &  0.57 &  0.25 &  0.38 &  0.56 &  0.21 &  0.48\\ 
314-037 &  7.85 & 4.83 & --0.170 & --0.141 &  0.94 &  0.45 &  6.59 &  5.07 & --0.12 &  1.31 &  3.28 &  1.57 &  4.70 &  1.70\\ 
... &  0.81 & 0.77 & 0.028 & 0.032 &  0.43 &  0.87 &  0.71 &  0.42 &  1.29 &  0.61 &  0.92 &  1.45 &  0.52 &  1.28\\ 
%316-04 &  4.29 & 2.46 & --0.070 & --0.063 &  0.89 &  4.26 & --0.78 &  1.51 &  0.96 & --0.14 &  3.55 &  3.84 &  3.96 &  4.34\\ 
% &  0.91 & 0.89 & 0.029 & 0.033 &  0.46 &  0.79 &  0.85 &  0.51 &  1.28 &  0.64 &  0.90 &  1.38 &  0.50 &  1.16\\ 
321-046 & 10.69 & 7.55 & --0.220 & --0.149 &  0.25 & --1.98 &  9.47 &  6.97 &  0.44 &  0.56 &  1.00 &  0.33 &  6.83 &  0.97\\ 
... &  0.61 & 0.54 & 0.022 & 0.024 &  0.32 &  0.63 &  0.45 &  0.26 &  0.89 &  0.42 &  0.66 &  0.98 &  0.32 &  0.85\\ 
322-049 &  7.88 & 5.77 & --0.160 & --0.106 &  0.01 & --1.68 &  7.09 &  5.53 & --0.06 &  0.00 &  0.93 &  0.83 &  5.04 &  1.23\\ 
... &  0.52 & 0.46 & 0.018 & 0.020 &  0.21 &  0.41 &  0.33 &  0.19 &  0.64 &  0.30 &  0.46 &  0.69 &  0.24 &  0.62\\ 
324-051 &  6.75 & 4.83 & --0.120 & --0.069 &  0.29 &  0.84 &  4.89 &  4.42 &  0.74 &  0.36 &  2.34 &  2.11 &  4.67 &  3.24\\ 
... &  0.46 & 0.40 & 0.016 & 0.017 &  0.13 &  0.24 &  0.23 &  0.13 &  0.45 &  0.18 &  0.27 &  0.42 &  0.14 &  0.38\\ 
327-053 &  5.85 & 4.40 & --0.080 & --0.030 &  0.05 & --1.03 &  5.56 &  4.58 &  0.37 &  0.52 &  0.72 &  1.04 &  3.76 &  1.78\\ 
... &  0.45 & 0.39 & 0.015 & 0.016 &  0.12 &  0.24 &  0.21 &  0.12 &  0.44 &  0.17 &  0.28 &  0.41 &  0.14 &  0.38\\ 
328-054 &  3.98 & 2.72 & --0.090 & --0.057 &  0.75 &  2.17 &  1.51 &  2.36 &  0.46 &  0.32 &  0.56 & --0.22 &  2.56 &  0.78\\ 
... &  0.70 & 0.66 & 0.022 & 0.025 &  0.32 &  0.60 &  0.57 &  0.34 &  0.93 &  0.44 &  0.68 &  0.99 &  0.35 &  0.81\\ 
%331-05 &  2.50 & 2.24 & -0.100 & -0.150 & -0.84 &  2.07 & -0.32 &  2.61 &  3.52 &  2.12 &  3.84 &  4.84 &  2.16 &  4.12\\ 
%331-05 &  0.94 & 0.89 & 0.029 & 0.033 &  0.53 &  0.86 &  0.86 &  0.49 &  1.26 &  0.62 &  0.95 &  1.45 &  0.57 &  1.26\\ 
337-068 &  2.66 & 2.29 & --0.060 & --0.030 &  0.60 &  3.73 & --0.02 &  1.79 &  2.22 &  0.58 &  2.10 &  1.60 &  3.21 &  2.65\\ 
... &  0.43 & 0.37 & 0.015 & 0.015 &  0.07 &  0.13 &  0.16 &  0.09 &  0.34 &  0.09 &  0.16 &  0.22 &  0.07 &  0.22\\ 
347-154 &  4.38 & 3.12 & --0.080 & --0.036 &  0.29 &  1.24 &  2.83 &  2.89 &  0.52 &  0.03 &  1.01 &  0.51 &  2.85 &  1.01\\ 
... &  0.51 & 0.45 & 0.017 & 0.018 &  0.17 &  0.31 &  0.29 &  0.18 &  0.53 &  0.22 &  0.34 &  0.50 &  0.17 &  0.43\\
350-162 &  3.50 & 2.69 & --0.050 & --0.013 &  0.35 &  2.55 &  1.11 &  2.12 &  1.29 &  0.21 &  1.20 &  0.89 &  2.78 &  1.59\\ 
... &  0.59 & 0.54 & 0.019 & 0.021 &  0.25 &  0.43 &  0.43 &  0.26 &  0.70 &  0.33 &  0.50 &  0.74 &  0.27 &  0.63\\ 
%354-18 &  3.36 & 2.92 & -0.060 & -0.014 &  0.02 &  1.25 &  2.45 &  2.64 &  0.78 & -0.02 &  0.28 &  1.70 &  2.33 &  0.75\\ 
%354-18 &  0.86 & 0.80 & 0.027 & 0.030 &  0.44 &  0.77 &  0.70 &  0.43 &  1.14 &  0.56 &  0.86 &  1.23 &  0.44 &  1.02\\ 
365-284 &  2.83 & 2.47 & --0.060 & --0.025 &  0.32 &  2.53 &  1.33 &  2.35 &  1.12 &  0.33 &  1.53 &  0.71 &  2.70 &  2.05\\ 
... &  0.52 & 0.46 & 0.017 & 0.018 &  0.17 &  0.30 &  0.30 &  0.18 &  0.53 &  0.22 &  0.34 &  0.50 &  0.17 &  0.43\\ 
380-313 &  9.60 & 6.46 & --0.190 & --0.123 &  0.41 & --0.17 &  8.20 &  6.32 & --0.08 &  0.37 &  1.84 &  0.61 &  6.05 &  2.74\\ 
... &  0.50 & 0.44 & 0.017 & 0.019 &  0.19 &  0.38 &  0.31 &  0.17 &  0.61 &  0.27 &  0.41 &  0.63 &  0.22 &  0.57\\ 
383-318 & --0.36 & 0.35 & 0.060 & 0.096 &  0.80 &  4.87 & --3.60 & --0.33 &  3.33 &  0.98 &  2.81 &  2.87 &  1.73 &  3.95\\ 
... &  0.48 & 0.43 & 0.016 & 0.017 &  0.13 &  0.22 &  0.26 &  0.16 &  0.43 &  0.16 &  0.25 &  0.37 &  0.13 &  0.33\\ 
393-330 & --0.03 & 0.69 & 0.000 & 0.024 &  0.72 &  3.86 & --2.55 & --0.23 &  2.99 &  0.49 &  1.76 &  1.02 &  1.88 &  2.82\\ 
... &  0.60 & 0.55 & 0.019 & 0.020 &  0.23 &  0.40 &  0.43 &  0.27 &  0.65 &  0.30 &  0.45 &  0.67 &  0.24 &  0.57\\ 
398-341 & --1.34 &--0.04 & 0.080 & 0.108 &  0.96 &  5.39 & --5.19 & --1.04 &  3.76 &  0.75 &  3.11 &  2.83 &  1.58 &  3.86\\ 
... &  0.63 & 0.58 & 0.019 & 0.021 &  0.23 &  0.39 &  0.46 &  0.29 &  0.64 &  0.30 &  0.43 &  0.65 &  0.24 &  0.55\\ 
401-344 &  4.51 & 3.55 & --0.090 & --0.054 &  0.01 &  1.49 &  2.69 &  2.75 &  0.21 & --0.01 &  0.40 & --0.26 &  2.82 &  0.59\\ 
... &  0.49 & 0.44 & 0.017 & 0.018 &  0.17 &  0.31 &  0.29 &  0.18 &  0.53 &  0.23 &  0.35 &  0.53 &  0.19 &  0.47\\ 
NB16 &  3.52 & 2.77 & --0.070 & --0.035 &  0.50 &  2.71 &  1.30 &  2.38 &  1.45 &  0.37 &  1.66 &  1.22 &  3.32 &  2.31\\ 
... &  0.43 & 0.37 & 0.015 & 0.015 &  0.07 &  0.14 &  0.16 &  0.09 &  0.35 &  0.10 &  0.17 &  0.25 &  0.08 &  0.25\\ 
NB67 &  3.29 & 2.71 & --0.070 & --0.033 &  0.48 &  2.73 &  1.20 &  2.26 &  1.44 &  0.29 &  1.76 &  1.09 &  3.29 &  2.38\\ 
... &  0.45 & 0.39 & 0.015 & 0.016 &  0.11 &  0.20 &  0.21 &  0.12 &  0.41 &  0.15 &  0.23 &  0.34 &  0.12 &  0.32\\ 
NB68 & --0.15 & 0.94 & --0.040 & --0.017 &  0.94 &  5.23 & --3.69 & --0.43 &  3.50 &  0.65 &  2.50 &  2.45 &  2.30 &  3.09\\ 
... &  0.45 & 0.39 & 0.015 & 0.016 &  0.10 &  0.18 &  0.21 &  0.13 &  0.38 &  0.13 &  0.21 &  0.30 &  0.10 &  0.28\\ 
NB74 & --2.90 &--0.58 & --0.020 & 0.001 &  1.53 &  6.44 & --6.92 & --2.24 &  5.29 &  0.67 &  2.47 &  2.12 &  1.22 &  2.95\\ 
... &  0.84 & 0.78 & 0.024 & 0.027 &  0.33 &  0.58 &  0.73 &  0.46 &  0.90 &  0.45 &  0.66 &  1.00 &  0.38 &  0.86\\ 
NB81 &  0.58 & 1.20 & --0.040 & --0.016 &  0.75 &  5.25 & --3.15 &  0.15 &  3.79 &  0.95 &  2.76 &  3.85 &  2.83 &  4.05\\ 
... &  0.46 & 0.41 & 0.015 & 0.016 &  0.12 &  0.20 &  0.24 &  0.15 &  0.41 &  0.15 &  0.24 &  0.35 &  0.12 &  0.33\\ 
NB83 &  2.61 & 2.27 & --0.060 & --0.023 &  0.66 &  3.56 &  0.22 &  1.96 &  1.97 &  0.49 &  2.09 &  1.62 &  3.21 &  2.63\\ 
... &  0.46 & 0.41 & 0.015 & 0.016 &  0.12 &  0.22 &  0.24 &  0.14 &  0.43 &  0.17 &  0.26 &  0.38 &  0.13 &  0.35\\ 
NB87 & --3.72 &--0.56 & 0.010 & 0.048 &  1.76 &  5.96 & --7.38 & --2.19 &  6.22 &  1.21 &  3.24 &  4.52 &  1.59 &  4.29\\ 
... &  0.50 & 0.43 & 0.016 & 0.017 &  0.12 &  0.22 &  0.28 &  0.17 &  0.42 &  0.16 &  0.25 &  0.36 &  0.13 &  0.33\\ 
NB89 &  0.70 & 1.31 & 0.050 & 0.094 &  0.69 &  4.04 & --1.61 &  0.84 &  2.85 &  0.88 &  2.65 &  2.11 &  2.02 &  3.28\\ 
... &  0.50 & 0.44 & 0.016 & 0.017 &  0.13 &  0.22 &  0.23 &  0.14 &  0.41 &  0.14 &  0.22 &  0.30 &  0.09 &  0.26\\ 
NB91 & --0.38 & 0.84 & --0.040 & --0.016 &  0.95 &  5.25 & --3.76 & --0.53 &  3.59 &  0.70 &  2.43 &  2.42 &  2.17 &  3.32\\ 
... &  0.47 & 0.41 & 0.015 & 0.016 &  0.11 &  0.20 &  0.23 &  0.15 &  0.40 &  0.14 &  0.23 &  0.33 &  0.12 &  0.30\\ 
\enddata

\end{deluxetable}
%\end{turnpage}

\clearpage

%\begin{turnpage}
\begin{deluxetable}{lrrrrrrrrrr} 
\renewcommand{\arraystretch}{.6} 
\rotate
\footnotesize
\tablecaption{Lick/IDS indices and uncertainties for M31 globular clusters} 
%\label{M31Indices1}
\tablenum{4A}
\tablewidth{0pt}
\tablehead{
\colhead{Name} & \colhead{Mg$_1$} & \colhead{Mg$_2$}  & \colhead{Mg $b$}  & \colhead{Fe5270} &
\colhead{Fe5335} & \colhead{Fe5406} & \colhead{Fe5709} & \colhead{Fe5782}  & \colhead{Na D}  & 
\colhead{TiO$_1$}\\
\colhead{} & \colhead{[mag]} &  \colhead{[mag]} & \colhead{[\AA]} & \colhead{[\AA]} &  
\colhead{[\AA]} & \colhead{[\AA]} &  \colhead{[\AA]} & \colhead{[\AA]} & \colhead{[\AA]} &  
\colhead{[mag]}\\
}
\startdata
126-184 & 0.014 &0.066 &  1.28 &  1.02 &  0.81 &  0.47 &  0.27 &  0.16 &  1.15 & 0.010\\ 
... & 0.015 &0.014 &  0.24 &  0.17 &  0.20 &  0.14 &  0.11 &  0.11 &  0.52 & 0.010\\ 
134-190 & 0.050 &0.129 &  2.22 &  1.79 &  1.59 &  0.99 &  0.61 &  0.41 &  1.75 & 0.022\\ 
... & 0.015 &0.014 &  0.24 &  0.17 &  0.20 &  0.15 &  0.11 &  0.11 &  0.52 & 0.010\\ 
158-213 & 0.074 &0.150 &  2.27 &  1.86 &  1.70 &  1.06 &  0.62 &  0.47 &  2.48 & 0.020\\ 
... & 0.015 &0.013 &  0.21 &  0.10 &  0.13 &  0.09 &  0.07 &  0.07 &  0.51 & 0.010\\ 
163-217 & 0.098 &0.242 &  4.01 &  2.60 &  2.44 &  1.57 &  0.83 &  0.76 &  4.30 & 0.038\\ 
... & 0.015 &0.013 &  0.19 &  0.07 &  0.09 &  0.06 &  0.04 &  0.05 &  0.50 & 0.010\\ 
222-277 & 0.050 &0.121 &  1.30 &  1.87 &  1.17 &  0.97 &  0.61 &  0.37 &  3.25 & 0.029\\ 
... & 0.016 &0.015 &  0.39 &  0.37 &  0.43 &  0.31 &  0.24 &  0.24 &  0.58 & 0.012\\ 
225-280 & 0.089 &0.207 &  3.21 &  2.31 &  2.03 &  1.31 &  0.79 &  0.65 &  3.10 & 0.027\\ 
... & 0.015 &0.013 &  0.19 &  0.07 &  0.09 &  0.06 &  0.05 &  0.05 &  0.50 & 0.010\\ 
234-290 & 0.055 &0.133 &  2.37 &  2.03 &  1.50 &  0.99 &  0.51 &  0.46 &  2.15 & 0.020\\ 
... & 0.015 &0.014 &  0.24 &  0.18 &  0.21 &  0.15 &  0.12 &  0.11 &  0.52 & 0.010\\ 
292-010 & 0.038 &0.073 &  0.97 &  0.95 &  1.14 &  0.15 &  0.33 &  0.19 &  1.32 & 0.006\\ 
... & 0.016 &0.016 &  0.40 &  0.39 &  0.45 &  0.34 &  0.26 &  0.25 &  0.60 & 0.013\\ 
301-022 & 0.022 &0.076 &  1.70 &  1.43 &  1.18 &  0.33 &  0.56 &  0.36 &  1.44 & 0.008\\ 
... & 0.016 &0.015 &  0.38 &  0.37 &  0.42 &  0.31 &  0.24 &  0.23 &  0.58 & 0.012\\ 
%302-02 & 0.030 &0.058 &  0.88 &  0.84 &  1.10 &  0.47 &  0.43 &  0.17 &  1.75 & 0.015\\ 
%... & 0.017 &0.017 &  0.46 &  0.48 &  0.55 &  0.41 &  0.32 &  0.30 &  0.63 & 0.013\\ 
304-028 & 0.041 &0.070 &  1.23 &  1.40 &  0.92 &  0.85 &  0.39 &  0.26 &  1.72 & 0.011\\ 
... & 0.016 &0.015 &  0.37 &  0.36 &  0.42 &  0.30 &  0.24 &  0.23 &  0.58 & 0.012\\ 
%305-D0 & 0.057 &0.118 &  1.77 &  1.66 &  0.95 &  1.35 &  1.01 &  1.12 &  2.85 & 0.008\\ 
%... & 0.020 &0.020 &  0.64 &  0.69 &  0.80 &  0.58 &  0.45 &  0.42 &  0.74 & 0.016\\ 
%307-03 & 0.079 &0.151 &  2.32 &  2.71 &  3.06 &  1.33 &  1.03 &  0.87 &  2.60 & 0.024\\ 
%... & 0.018 &0.018 &  0.50 &  0.51 &  0.58 &  0.44 &  0.34 &  0.33 &  0.66 & 0.014\\ 
310-032 & 0.035 &0.055 &  0.92 &  1.26 &  0.84 &  0.44 &  0.25 &  0.02 &  1.57 & 0.017\\ 
... & 0.016 &0.015 &  0.36 &  0.35 &  0.40 &  0.29 &  0.23 &  0.22 &  0.57 & 0.012\\ 
313-036 & 0.060 &0.140 &  2.17 &  1.69 &  1.45 &  0.97 &  0.66 &  0.61 &  2.80 & 0.024\\ 
... & 0.015 &0.014 &  0.28 &  0.23 &  0.26 &  0.19 &  0.15 &  0.14 &  0.53 & 0.011\\ 
314-037 & 0.055 &0.122 &  2.85 &  1.56 &  1.89 &  0.82 &  0.80 &  0.31 &  2.73 & --0.004\\ 
... & 0.020 &0.020 &  0.61 &  0.69 &  0.79 &  0.59 &  0.47 &  0.47 &  0.77 & 0.017\\ 
%316-04 & 0.068 &0.129 &  1.57 &  3.84 &  1.83 &  1.77 &  0.88 &  0.99 &  1.11 & 0.034\\ 
%... & 0.019 &0.019 &  0.59 &  0.59 &  0.70 &  0.50 &  0.40 &  0.38 &  0.71 & 0.016\\ 
321-046 & 0.028 &0.052 &  0.94 &  0.80 &  0.73 &  0.20 &  0.45 &  0.35 &  1.37 & 0.007\\ 
... & 0.017 &0.016 &  0.43 &  0.45 &  0.52 &  0.38 &  0.30 &  0.29 &  0.62 & 0.013\\ 
322-049 & 0.038 &0.048 &  0.35 &  0.63 &  0.67 &  0.45 &  0.46 &  0.28 &  1.68 & 0.014\\ 
... & 0.016 &0.015 &  0.34 &  0.33 &  0.38 &  0.28 &  0.23 &  0.22 &  0.58 & 0.012\\ 
324-051 & 0.045 &0.085 &  1.57 &  1.66 &  1.46 &  0.71 &  0.55 &  0.31 &  0.71 & 0.019\\ 
... & 0.015 &0.014 &  0.24 &  0.19 &  0.22 &  0.16 &  0.13 &  0.13 &  0.53 & 0.011\\ 
327-053 & 0.050 &0.077 &  0.59 &  0.83 &  1.10 &  0.72 &  0.42 &  0.60 &  2.67 & 0.028\\ 
... & 0.015 &0.014 &  0.25 &  0.19 &  0.22 &  0.16 &  0.13 &  0.12 &  0.52 & 0.011\\ 
328-054 & 0.033 &0.068 &  0.19 &  0.89 &  0.49 &  0.40 &  0.04 &  0.44 &  1.88 & 0.002\\ 
... & 0.017 &0.016 &  0.42 &  0.41 &  0.48 &  0.35 &  0.27 &  0.25 &  0.59 & 0.013\\ 
%331-05 & 0.075 &0.099 &  2.06 &  2.79 &  1.29 & --0.01 &  1.77 &  0.06 &  4.11 & 0.030\\ 
%... & 0.020 &0.020 &  0.63 &  0.68 &  0.79 &  0.61 &  0.45 &  0.46 &  0.74 & 0.017\\ 
337-068 & 0.013 &0.084 &  1.86 &  1.48 &  1.11 &  0.64 &  0.44 &  0.26 &  1.21 & 0.006\\ 
... & 0.015 &0.013 &  0.19 &  0.07 &  0.09 &  0.06 &  0.05 &  0.05 &  0.50 & 0.010\\ 
347-154 & 0.020 &0.044 &  0.76 &  0.51 &  0.49 &  0.37 &  0.12 &  0.11 &  1.15 & 0.002\\ 
... & 0.015 &0.014 &  0.26 &  0.21 &  0.24 &  0.17 &  0.14 &  0.13 &  0.53 & 0.011\\ 
350-162 & 0.034 &0.075 &  1.13 &  0.98 &  0.81 &  0.33 &  0.39 &  0.19 &  1.00 & 0.006\\ 
... & 0.016 &0.015 &  0.34 &  0.32 &  0.37 &  0.27 &  0.21 &  0.20 &  0.56 & 0.012\\ 
%354-18 & 0.056 &0.073 &  0.57 &  0.89 &  0.63 &  0.29 & --0.01 &  0.22 &  1.55 & --0.006\\ 
%... & 0.018 &0.017 &  0.50 &  0.51 &  0.59 &  0.44 &  0.34 &  0.32 &  0.64 & 0.013\\ 
365-284 & 0.040 &0.080 &  1.37 &  1.33 &  1.00 &  0.50 &  0.43 &  0.31 &  2.54 & 0.014\\ 
... & 0.015 &0.014 &  0.26 &  0.20 &  0.24 &  0.18 &  0.14 &  0.13 &  0.53 & 0.011\\ 
380-313 & 0.038 &0.080 &  1.19 &  1.86 &  1.36 &  0.64 &  0.50 &  0.42 &  1.96 & 0.026\\ 
... & 0.016 &0.014 &  0.32 &  0.29 &  0.34 &  0.26 &  0.21 &  0.20 &  0.56 & 0.012\\ 
383-318 & 0.070 &0.183 &  3.03 &  2.12 &  1.79 &  1.21 &  0.63 &  0.51 &  2.38 & 0.028\\ 
... & 0.015 &0.013 &  0.22 &  0.14 &  0.17 &  0.12 &  0.09 &  0.09 &  0.51 & 0.010\\ 
393-330 & 0.069 &0.122 &  1.49 &  1.93 &  1.58 &  0.70 &  0.59 &  0.48 &  1.92 & 0.018\\ 
... & 0.016 &0.014 &  0.32 &  0.28 &  0.33 &  0.24 &  0.19 &  0.18 &  0.55 & 0.011\\ 
398-341 & 0.086 &0.182 &  3.17 &  2.31 &  1.71 &  1.07 &  0.65 &  0.45 &  2.49 & 0.028\\ 
... & 0.016 &0.014 &  0.30 &  0.27 &  0.31 &  0.23 &  0.18 &  0.17 &  0.54 & 0.011\\ 
401-344 & 0.034 &0.047 &  0.53 &  0.63 &  0.53 & --0.03 & --0.01 &  0.06 &  0.93 & 0.005\\ 
... & 0.015 &0.014 &  0.27 &  0.23 &  0.27 &  0.20 &  0.16 &  0.15 &  0.54 & 0.011\\ 
NB16 & 0.023 &0.086 &  1.61 &  1.18 &  0.97 &  0.50 &  0.39 &  0.20 &  1.16 & 0.006\\ 
... & 0.015 &0.013 &  0.20 &  0.09 &  0.11 &  0.08 &  0.06 &  0.07 &  0.51 & 0.010\\ 
NB67 & 0.023 &0.081 &  1.65 &  1.12 &  0.79 &  0.47 &  0.32 &  0.23 &  1.14 & 0.003\\ 
... & 0.015 &0.013 &  0.22 &  0.14 &  0.17 &  0.12 &  0.09 &  0.09 &  0.51 & 0.010\\ 
NB68 & 0.033 &0.145 &  3.22 &  1.98 &  1.59 &  0.90 &  0.59 &  0.40 &  1.51 & 0.005\\ 
... & 0.015 &0.013 &  0.21 &  0.12 &  0.14 &  0.10 &  0.08 &  0.08 &  0.51 & 0.010\\ 
NB74 & 0.080 &0.239 &  5.17 &  1.82 &  1.83 &  0.97 &  0.72 &  0.34 &  3.10 & --0.003\\ 
... & 0.017 &0.016 &  0.42 &  0.45 &  0.52 &  0.39 &  0.31 &  0.30 &  0.62 & 0.013\\ 
NB81 & 0.041 &0.142 &  2.72 &  2.19 &  1.75 &  0.90 &  0.78 &  0.48 &  1.51 & 0.002\\ 
... & 0.015 &0.013 &  0.22 &  0.14 &  0.17 &  0.12 &  0.10 &  0.10 &  0.51 & 0.010\\ 
NB83 & 0.023 &0.087 &  1.79 &  1.44 &  1.09 &  0.42 &  0.38 &  0.26 &  1.18 & 0.004\\ 
... & 0.015 &0.014 &  0.23 &  0.16 &  0.19 &  0.14 &  0.11 &  0.11 &  0.52 & 0.010\\ 
NB87 & 0.089 &0.277 &  5.19 &  3.25 &  2.85 &  1.88 &  0.98 &  0.75 &  2.99 & 0.006\\ 
... & 0.015 &0.013 &  0.22 &  0.14 &  0.17 &  0.12 &  0.09 &  0.09 &  0.51 & 0.010\\ 
NB89 & 0.054 &0.143 &  2.43 &  1.91 &  1.63 &  1.02 &  0.59 &  0.50 &  3.08 & 0.019\\ 
... & 0.015 &0.013 &  0.20 &  0.09 &  0.11 &  0.07 &  0.06 &  0.06 &  0.50 & 0.010\\ 
NB91 & 0.034 &0.145 &  3.19 &  1.96 &  1.76 &  0.98 &  0.65 &  0.37 &  1.79 & 0.004\\ 
... & 0.015 &0.013 &  0.21 &  0.13 &  0.16 &  0.11 &  0.09 &  0.09 &  0.51 & 0.010\\ 
\enddata
\end{deluxetable}

\clearpage%%%%Galactic Lick Offsets

\begin{deluxetable}{lllcccc}
\footnotesize
\tablecaption{Linear fit coefficients, rms of linear fit, 
mean difference and its standard deviation between the CBR98
Galactic cluster data and Trager et al. (1998) Galactic clusters.
Lick indices for H$\delta$ and H$\gamma$ are taken from 
Kuntschner et al. (2003). 
N signifies number of clusters with common indices. 
Indices measured using the Trager et al. (1998) and Worthey \& Ottaviani (1997)
Lick index definitions. \label{CBR98Offsets}}
\tablewidth{0pt}
\tablehead{
\colhead{Index} & \colhead{$a$} & \colhead{$b$} & \colhead{rms} & 
\colhead{Lick--CBR98}   & \colhead{$\sigma_{\rm rms}$} & \colhead{N}\\
}
\startdata
H$\delta_{\rm A}$	& ... & ... & ... & ... & ... & ... \\  
H$\delta_{\rm F}$	&0.6108 & --1.3760 & ... & 0.2210 & 1.1430 & 2$^a$ \\  
CN$_1$ 		 	&0.0725 & --1.0540 & ... & 0.0210 & 0.0537 & 2$^a$ \\  
CN$_2$		 	&0.0052 & 0.8462 & ... & 0.0020 & 0.0057 & 2$^a$ \\   
Ca4227		 	&0.2373 & 0.4451 & 0.0631 & 0.1067 & 0.3919 & 4$^b$ \\   
G4300		 	&0.2390 & 1.0550 & 0.4266 & --0.4015 & 0.5039 & 4$^b$ \\  
H$\gamma_{\rm A}$	&--1.1790 & 2.4030 & 0.1657 & 0.5383 & 2.3030 & 3 \\  
H$\gamma_{\rm F}$	&--0.1537 & 0.9790 & 0.7687 & 0.1710 & 0.8882 & 4 \\  
Fe4383		 	&0.4878 & 0.5728 & 0.3304 & 0.6123 & 1.0930 & 4$^b$ \\   
Ca4455		 	&--0.1364 & 0.5420 & 0.0373 & 0.5220 & 0.2441 & 5 \\  
Fe4531		 	&--0.3698 & 1.0900 & 0.0702 & 0.1987 & 0.1172 & 4$^b$ \\  
C$_2$4668	 	&0.2719 & 0.7662 & 0.4248 & 0.0274 & 0.5502 & 5$^b$ \\   
H$\beta$	 	&0.2606 & 0.8592 & 0.1256 & 0.0077 & 0.1570 & 6 \\   
Fe5015		 	&--0.3993 & 1.0290 & 0.2905 & 0.3152 & 0.3202 & 6 \\  
Mg$_1$		 	&0.0002 & 1.1940 & 0.0171 & --0.0068 & 0.0192 & 6 \\  
Mg$_2$		 	&0.0091 & 0.9802 & 0.0010 & --0.0070 & 0.0017 & 6 \\  
Mg $b$		 	&--0.0859 & 1.0300 & 0.1239 & 0.0275 & 0.1392 & 6 \\  
Fe5270		 	&0.0543 & 0.9417 & 0.2217 & 0.0348 & 0.2461 & 6 \\   
Fe5335		 	&0.3537 & 0.8841 & 0.2172 & --0.2305 & 0.2492 & 6 \\  
Fe5406		 	&--0.1683 & 1.0310 & 0.0463 & 0.1390 & 0.0526 & 6 \\  
Fe5709		 	&0.1194 & 0.7490 & 0.1511 & 0.0106 & 0.1934 & 5$^b$ \\   
Fe5782		 	&0.0567 & 0.8221 & 0.1144 & 0.0180 & 0.1345 & 6 \\   
NaD		 	&0.1374 & 0.8028 & 0.1840 & 0.2898 & 0.2990 & 6 \\   
TiO$_1$		 	&0.0078 & 0.5477 & 0.0068 & 0.0005 & 0.0112 & 6 \\   
TiO$_2$		 	&--0.0102 & 1.3910 & 0.0154 & --0.0052 & 0.0199 & 4 \\ 
\enddata
%% Text for table notes should follow after the \enddata but before
%% the \end{deluxetable}. Make sure there is at least one \tablenotemark
%% in the table for each \tablenotetext.

\tablenotetext{a}{No rms calculated for N$<$3.}
\tablenotetext{b}{Index measurement for NGC~6171 excluded from fit.}

\end{deluxetable}

\clearpage%%%%M31 Lick Offsets

\begin{deluxetable}{lllcccc}
\footnotesize
\tablecaption{Linear fit coefficients, rms of linear fit, 
mean difference and its standard deviation between the three
clusters in common between the LRIS M31 globular cluster data and 
Trager et al. (1998). Lick indices for H$\delta$ and H$\gamma$ are 
taken from Kuntschner et al. (2003).
Lick index definitions are given in
Trager et al. (1998) and Worthey \& Ottaviani (1997)
\label{M31Offsets}}
\tablewidth{0pt}
\tablehead{
\colhead{Index} & \colhead{$a$} & \colhead{$b$} & \colhead{rms} & 
\colhead{Lick--M31}   & \colhead{$\sigma_{\rm rms}$} & \colhead{N}\\
}
\startdata
H$\delta_{\rm A}$	& --1.5190 & --0.4680 & ... & --1.2330 & 1.0690 & 2$^a$ \\ 
H$\delta_{\rm F}$	& 0.6042 & 0.2634 & 0.0143 & --0.0370 & 0.7143 & 3 \\   
CN$_1$ 		  	& 0.0308 & 1.0480 & 0.0048 & --0.0337 & 0.0065 & 3 \\   
CN$_2$		 	& 0.0334 & 0.9869 & 0.0063 & --0.0320 & 0.0078 & 3 \\   
Ca4227		 	& 0.4951 & 0.4415 & 0.1514 & --0.0967 & 0.2211 & 3 \\   
G4300		   	& --0.1284 & 1.0120 & 0.1967 & 0.0733 & 0.2410 & 3 \\
H$\gamma_{\rm A}$	& 5.3500 & 1.9070 & ... & --0.6725 & 0.5197 & 2$^a$ \\   	
H$\gamma_{\rm F}$	& --0.2218 & 0.6410 & 0.1045 & 0.1500 & 0.3576 & 3 \\   
Fe4383		 	& 4.5760 & --0.2190 & 0.6566 & 0.0233 & 0.9420 & 3 \\   
Ca4455		 	& --0.3685 & 1.0790 & 0.0962 & 0.2720 & 0.1196 & 3 \\   
Fe4531		 	& 1.0450 & 0.6931 & 0.0273 & --0.2370 & 0.2176 & 3 \\   
C$_2$4668	        & 1.8400 & 0.4738 & 0.2666 & --0.5613 & 1.1520 & 3 \\    
H$\beta$	 	& 1.4620 & 0.1640 & 0.0273 & 0.0150 & 0.2247 & 3 \\     
Fe5015		   	& 0.9523 & 0.7859 & 0.2682 & --0.0973 & 0.3823 & 3 \\   
Mg$_1$		 	& 0.0646 & 0.4040 & 0.0066 & --0.0317 & 0.0156 & 3 \\   
Mg$_2$		 	& 0.0236 & 0.9728 & 0.0031 & --0.0187 & 0.0040 & 3 \\    
Mg $b$		 	& --0.5430 & 1.2040 & 0.0946 & --0.0840 & 0.1863 & 3 \\ 
Fe5270		 	& --0.0404 & 0.9489 & 0.1330 & 0.1640 & 0.1639 & 3 \\   
Fe5335		 	& 0.4890 & 0.8229 & 0.1629 & --0.1513 & 0.2104 & 3 \\    
Fe5406		 	& --0.5510 & 1.6110 & 0.1155 & --0.1560 & 0.1633 & 3 \\ 
Fe5709		 	& 0.4219 & 0.3616 & 0.0688 & 0.1527 & 0.1622 & 3 \\    	
Fe5782		 	& 0.4351 & 0.2569 & 0.0846 & 0.1173 & 0.3092 & 3 \\    	
NaD		 	& --0.5590 & 1.1620 & 0.1179 & 0.0210 & 0.1932 & 3 \\   
TiO$_1$		   	& 0.0186 & 0.4208 & 0.0032 & --0.0053 & 0.0119 & 3 \\   
TiO$_2$		 	& ... & ... & ... & ... & ... & ... \\   		
\enddata
%% Text for table notes should follow after the \enddata but before
%% the \end{deluxetable}. Make sure there is at least one \tablenotemark
%% in the table for each \tablenotetext.

\tablenotetext{a}{No rms calculated for N$<$3.}
\end{deluxetable}

\clearpage

%%%%LMC Correlations

\begin{deluxetable}{llllc}
\footnotesize
\tablecaption{Results of cross-correlation between young M31 
clusters and the LMC templates of Beasley et al. 2002. 
LMC \#1..\#3 denote the first, second and third best
template matches for the M31 cluster in question.\label{CrossCorrTab}}
\tablewidth{0pt}
\tablehead{
\colhead{ID} & \colhead{LMC \#1} & \colhead{LMC \#2} & \colhead{LMC \#3} &
\colhead{Age Range (Gyr)} \\
}
\startdata
327-053 & SL106 (I) & NGC~2002(I) & NGC~1863(I) & 0.01--0.03 \\
322-049 & NGC~2031 (III) & NGC~1735 (II) & NGC~1815 (I) & 0.01--0.20 \\
380-313 & NGC~1735 (II) & NGC~2031 (III) & NGC~1755 (II) & 0.01--0.20 \\
321-046 & NGC~2031 (III) & NGC~1735 (II) & NGC~1774 (II) & 0.01--0.20 \\
324-051 & NGC~1940 (II) & NGC~2041 (II) & NGC~2127( III) & 0.01--0.20 \\
314-037 & NGC~2127 (III) & NGC~2041 (II) & NGC~1878 (IVA) & 0.03--0.40 \\
222-277 & NGC~1878 (IVA) & NGC2127 (III) & NGC~1806 (V) & 0.07--2.00\\
292-010 & NGC~1852 (V) & NGC~1751 (V) & NGC~1718 (VI) & 0.80--2.0 \\ 
\enddata
%% Text for table notes should follow after the \enddata but before
%% the \end{deluxetable}. Make sure there is at least one \tablenotemark
%% in the table for each \tablenotetext.

\tablecomments{Roman numerals in parenthesis denote SWB-type, originally
defined by Searle, Wilkinson \& Bagnuolo (1980).
Age--SWB calibration from Bica et al. (1992)}  

\end{deluxetable}

\clearpage

%%%%Chi Squared

\begin{deluxetable}{lcccccc}
\footnotesize
\tablecaption{Best matches to the BC03 stellar population models
using a $\chi^2$ minimization of our measured Lick indices for
the young M31 clusters. N refers to the number of Lick indices
utilised in the fitting procedure.
Re-measured velocities for the clusters
using new templates are also given.\label{ChiTab}}
\tablewidth{0pt}
\tablehead{
\colhead{ID} & \colhead{[Fe/H]} & \colhead{Age(Gyr)} &
\colhead{N} & \colhead{$M/L_V$} & \colhead{Mass($\Msun$)} & \colhead{Velocity (kms$^{-1}$)} \\
}
\startdata
327-053 & 0.300 $\pm$   0.750   & 0.080 $\pm$ 0.929 &  18 & 0.092 & $2.1\times10^4$ & --542$\pm$56\\
322-049 & --0.200 $\pm$   0.200  &  0.102 $\pm$   0.470 &  24 & 0.104 & $7.0\times10^3$ & --576$\pm$21\\
380-313 &  0.150 $\pm$   0.100  &  0.454 $\pm$   0.133 &    24 & 0.250&$4.1\times10^4$ & --115$\pm$24\\
321-046 &  0.000 $\pm$   0.600  &  0.286 $\pm$   0.280 &   24 & 0.178&$1.4\times10^4$ & --513$\pm$32\\
324-051 &  --0.050 $\pm$    0.400&    0.839 $\pm$    0.167 &    23 & 0.378&$6.5\times10^4$ & --266$\pm$38\\
314-037 &  0.350 $\pm$   0.250  &  0.509 $\pm$   0.590  &   22 & 0.268&$2.3\times10^4$& --484$\pm$54\\
222-277 &  0.300 $\pm$   0.600  &  0.719 $\pm$   0.716  &    22 & 0.346& $4.5\times10^4$& --320$\pm$55\\
292-010 & --1.350 $\pm$   0.550  &  2.748 $\pm$   1.151  &    22 & 1.110& $1.7\times10^5$& --392$\pm$56\\
\enddata
%% Text for table notes should follow after the \enddata but before
%% the \end{deluxetable}. Make sure there is at least one \tablenotemark
%% in the table for each \tablenotetext.

\end{deluxetable}

\end{document}